\documentclass[12pt,a4paper]{article}
\usepackage{ifthen} 
\newboolean{pdflatex}
\setboolean{pdflatex}{true} 

\newboolean{articletitles}
\setboolean{articletitles}{true} 

\newboolean{uprightparticles}
\setboolean{uprightparticles}{false} 
\newboolean{prl}
\setboolean{prl}{false}
\newboolean{wordcount}
\setboolean{wordcount}{false}

\def\paperauthors{LHCb collaboration} 
\def\paperasciititle{Measurement of the time-integrated CP asymmetry in D0 -> K K decays} \def\papertitle{Measurement of the time-integrated $\CP$ asymmetry in $\Dz\to\Km\Kp$ decays} 

\def\paperkeywords{{High Energy Physics}, {LHCb}} \def\papercopyright{\the\year\ CERN for the benefit of the LHCb collaboration} 
\def\paperlicence{CC BY 4.0 licence}
\def\paperlicenceurl{https://creativecommons.org/licenses/by/4.0/}

\usepackage[top=1in, bottom=1.25in, left=1in, right=1in]{geometry}

\usepackage{microtype}
\usepackage{lineno}  
\usepackage{xspace} 
\usepackage{caption} 

\usepackage{graphicx}  
\usepackage{color}
\usepackage{colortbl}
\graphicspath{{./figs/}} 

\usepackage{amsmath} 
\usepackage{amssymb}
\usepackage{amsfonts}
\usepackage{upgreek} 

\newcommand*\patchAmsMathEnvironmentForLineno[1]{%
\expandafter\let\csname old#1\expandafter\endcsname\csname #1\endcsname
\expandafter\let\csname oldend#1\expandafter\endcsname\csname
end#1\endcsname
 \renewenvironment{#1}%
   {\linenomath\csname old#1\endcsname}%
   {\csname oldend#1\endcsname\endlinenomath}%
}
\newcommand*\patchBothAmsMathEnvironmentsForLineno[1]{%
  \patchAmsMathEnvironmentForLineno{#1}%
  \patchAmsMathEnvironmentForLineno{#1*}%
}
\AtBeginDocument{%
\patchBothAmsMathEnvironmentsForLineno{equation}%
\patchBothAmsMathEnvironmentsForLineno{align}%
\patchBothAmsMathEnvironmentsForLineno{flalign}%
\patchBothAmsMathEnvironmentsForLineno{alignat}%
\patchBothAmsMathEnvironmentsForLineno{gather}%
\patchBothAmsMathEnvironmentsForLineno{multline}%
\patchBothAmsMathEnvironmentsForLineno{eqnarray}%
}

\usepackage{hyperxmp}

\usepackage[pdftex,
            pdfauthor={\paperauthors},
            pdftitle={\paperasciititle},
            pdfkeywords={\paperkeywords},
            pdfcopyright={Copyright (C) \papercopyright},
            pdflicenseurl={\paperlicenceurl}]{hyperref}

\usepackage[colorinlistoftodos,textsize=scriptsize]{todonotes}

\usepackage[bottom,flushmargin,hang,multiple]{footmisc}

\usepackage[all]{hypcap} 

\usepackage{xspace} 
\usepackage{upgreek}

\def\lhcb   {\mbox{LHCb}\xspace}

\def\MagUp {\mbox{\em Mag\kern -0.05em Up}\xspace}

\ifthenelse{\boolean{uprightparticles}}%
{

 \def\Ppi         {\ensuremath{\uppi}\xspace}

 \def\PDelta      {\ensuremath{\Delta}\xspace}                 
 \def\PXi         {\ensuremath{\Xi}\xspace}                 
 \def\PLambda     {\ensuremath{\Lambda}\xspace}                 
 \def\PSigma      {\ensuremath{\Sigma}\xspace}                 
 \def\POmega      {\ensuremath{\Omega}\xspace}                 
 \def\PUpsilon    {\ensuremath{\Upsilon}\xspace}
 \let\oldPi\Pi
 \def\PPi         {\ensuremath{\oldPi}\xspace}

 \def\PB      {\ensuremath{\mathrm{B}}\xspace}                 
                  
 \def\PD      {\ensuremath{\mathrm{D}}\xspace}

 \def\PK      {\ensuremath{\mathrm{K}}\xspace}

 \def\Pb      {\ensuremath{\mathrm{b}}\xspace}                 
 \def\Pc      {\ensuremath{\mathrm{c}}\xspace}                 
                  
 \def\Pe      {\ensuremath{\mathrm{e}}\xspace}

 \def\Pi      {\ensuremath{\mathrm{i}}\xspace}

 \def\Ps      {\ensuremath{\mathrm{s}}\xspace}

 \def\thebaroffset{0.0em}
}
{

 \def\Ppi         {\ensuremath{\pi}\xspace}

 \mathchardef\PDelta="7101
 \mathchardef\PXi="7104
 \mathchardef\PLambda="7103
 \mathchardef\PSigma="7106
 \mathchardef\POmega="710A
 \mathchardef\PUpsilon="7107
 \mathchardef\PPi="7105
                  
 \def\PB      {\ensuremath{B}\xspace}                 
                  
 \def\PD      {\ensuremath{D}\xspace}

 \def\PK      {\ensuremath{K}\xspace}

 \def\Pb      {\ensuremath{b}\xspace}                 
 \def\Pc      {\ensuremath{c}\xspace}                 
                  
 \def\Pe      {\ensuremath{e}\xspace}

 \def\Pi      {\ensuremath{i}\xspace}

 \def\Ps      {\ensuremath{s}\xspace}

 \def\thebaroffset{0.18em}
}
\newcommand{\offsetoverline}[2][\thebaroffset]{\kern #1\overline{\kern -#1 #2}}%

\makeatletter
\ifcase \@ptsize \relax
  \newcommand{\miniscule}{\@setfontsize\miniscule{4}{5}}
\or
  \newcommand{\miniscule}{\@setfontsize\miniscule{5}{6}}
\or
  \newcommand{\miniscule}{\@setfontsize\miniscule{5}{6}}
\fi
\makeatother

\DeclareRobustCommand{\optbar}[1]{\shortstack{{\miniscule (\rule[.5ex]{1.25em}{.18mm})}
  \\ [-.7ex] $#1$}}

\def\ep         {{\ensuremath{\Pe^+}}\xspace}

\def\squark    {{\ensuremath{\Ps}}\xspace}

\def\cquark    {{\ensuremath{\Pc}}\xspace}

\def\bquark    {{\ensuremath{\Pb}}\xspace}

\def\pion   {{\ensuremath{\Ppi}}\xspace}
\def\piz    {{\ensuremath{\pion^0}}\xspace}
\def\pip    {{\ensuremath{\pion^+}}\xspace}
\def\pim    {{\ensuremath{\pion^-}}\xspace}

\def\kaon    {{\ensuremath{\PK}}\xspace}
\def\Kbar    {{\ensuremath{\offsetoverline{\PK}}}\xspace}

\def\KorKbar {\kern \thebaroffset\optbar{\kern -\thebaroffset \PK}{}\xspace}
\def\Kz      {{\ensuremath{\kaon^0}}\xspace}
\def\Kzb     {{\ensuremath{\Kbar{}^0}}\xspace}
\def\Kp      {{\ensuremath{\kaon^+}}\xspace}
\def\Km      {{\ensuremath{\kaon^-}}\xspace}

\def\KS      {{\ensuremath{\kaon^0_{\mathrm{S}}}}\xspace}

\def\Dbar    {{\ensuremath{\offsetoverline{\PD}}}\xspace}
\def\D       {{\ensuremath{\PD}}\xspace}

\def\DorDbar {\kern \thebaroffset\optbar{\kern -\thebaroffset \PD}\xspace}
\def\Dz      {{\ensuremath{\D^0}}\xspace}
\def\Dzb     {{\ensuremath{\Dbar{}^0}}\xspace}
\def\Dp      {{\ensuremath{\D^+}}\xspace}
\def\Dm      {{\ensuremath{\D^-}}\xspace}

\def\DpDm    {\ensuremath{\Dp {\kern -0.16em \Dm}}\xspace}
\def\Dstar   {{\ensuremath{\D^*}}\xspace}

\def\Dstarp  {{\ensuremath{\D^{*+}}}\xspace}
\def\Dstarm  {{\ensuremath{\D^{*-}}}\xspace}

\def\Ds      {{\ensuremath{\D^+_\squark}}\xspace}
\def\Dsp     {{\ensuremath{\D^+_\squark}}\xspace}

\def\DporDsp {{\ensuremath{\D_{(\squark)}^+}}\xspace}

\def\B       {{\ensuremath{\PB}}\xspace}

\def\BorBbar {\kern \thebaroffset\optbar{\kern -\thebaroffset \PB}\xspace}

\def\Bd      {{\ensuremath{\B^0}}\xspace}

\def\BdorBdbar {\kern \thebaroffset\optbar{\kern -\thebaroffset \Bd}\xspace}

\def\Bs      {{\ensuremath{\B^0_\squark}}\xspace}

\def\BsorBsbar {\kern \thebaroffset\optbar{\kern -\thebaroffset \Bs}\xspace}

\def\Y#1S{\ensuremath{\PUpsilon{(#1S)}}\xspace}

\def\LorLbar     {\kern \thebaroffset\optbar{\kern -\thebaroffset \PLambda}\xspace}

\def\to                 {\ensuremath{\rightarrow}\xspace}

\def\CP                {{\ensuremath{C\!P}}\xspace}

\newcommand{\ACP}{{\ensuremath{{\mathcal{A}}^{\CP}}}\xspace}

\def\AT#1     {\ensuremath{A_{\mathrm{T}}^{#1}}\xspace}           

\def\C#1      {\ensuremath{\mathcal{C}_{#1}}\xspace}                       
\def\Cp#1     {\ensuremath{\mathcal{C}_{#1}^{'}}\xspace}                    
\def\Ceff#1   {\ensuremath{\mathcal{C}_{#1}^{\mathrm{(eff)}}}\xspace}        
\def\Cpeff#1  {\ensuremath{\mathcal{C}_{#1}^{'\mathrm{(eff)}}}\xspace}       
\def\Ope#1    {\ensuremath{\mathcal{O}_{#1}}\xspace}                       
\def\Opep#1   {\ensuremath{\mathcal{O}_{#1}^{'}}\xspace}                    

\newcommand{\aunit}[1]{\ensuremath{\text{\,#1}}}       

\newcommand{\tev}{\aunit{Te\kern -0.1em V}\xspace}
\newcommand{\gev}{\aunit{Ge\kern -0.1em V}\xspace}
\newcommand{\mev}{\aunit{Me\kern -0.1em V}\xspace}
\newcommand{\kev}{\aunit{ke\kern -0.1em V}\xspace}
\newcommand{\ev}{\aunit{e\kern -0.1em V}\xspace}
 
\newcommand{\mevc}{\ensuremath{\aunit{Me\kern -0.1em V\!/}c}\xspace}
\newcommand{\gevc}{\ensuremath{\aunit{Ge\kern -0.1em V\!/}c}\xspace}
\newcommand{\mevcc}{\ensuremath{\aunit{Me\kern -0.1em V\!/}c^2}\xspace}
\newcommand{\gevcc}{\ensuremath{\aunit{Ge\kern -0.1em V\!/}c^2}\xspace}

\def\mm   {\aunit{mm}\xspace}

\def\fb   {\ensuremath{\aunit{fb}}\xspace}
\def\invfb   {\ensuremath{\fb^{-1}}\xspace}

\newcommand{\stat}{\aunit{(stat)}\xspace}
\newcommand{\syst}{\aunit{(syst)}\xspace}

\newcommand{\chisq}{\ensuremath{\chi^2}\xspace}

\def\deriv {\ensuremath{\mathrm{d}}}

\def\gsim{{~\raise.15em\hbox{$>$}\kern-.85em
          \lower.35em\hbox{$\sim$}~}\xspace}
\def\lsim{{~\raise.15em\hbox{$<$}\kern-.85em
          \lower.35em\hbox{$\sim$}~}\xspace}

\def\pt         {\ensuremath{p_{\mathrm{T}}}\xspace}

\def\tell1  {TELL1\xspace}
\def\ukl1   {UKL1\xspace}

\newcommand{\ie}{\mbox{\itshape i.e.}\xspace}

\newcommand{\lhcborcid}[1]{\href{https://orcid.org/#1}{\hspace*{0.1em}\raisebox{-0.45ex}{\includegraphics[width=1em]{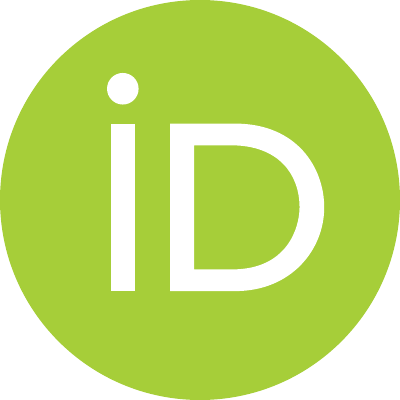}}}}

\usepackage{cite} 
\usepackage{mciteplus}

\newcommand{\Ameas}{\ensuremath{A}\xspace}

\newcommand{\Adet}{\ensuremath{A_{\rm D}}\xspace}
\newcommand{\AP}{\ensuremath{A_{\rm P}}\xspace}
\newcommand{\Aprod}{\AP}

\newcommand{\Acp}{\ensuremath{\ACP}\xspace}
\newcommand{\KK}{\ensuremath{\Km\Kp}\xspace}
\newcommand{\pipi}{\ensuremath{\pim\pip}\xspace}

\newcommand{\ACPKK}{\ensuremath{\ACP(\KK)}\xspace}
\newcommand{\acpkk}{\ensuremath{\ACP(\KK)}\xspace}
\newcommand{\ACPpipi}{\ensuremath{\ACP(\pim\pip)}\xspace}

\def \DACP {{\ensuremath{\Delta A_\CP\xspace}}}

\newcommand{\pitag}{\ensuremath{\pi^+_{\rm tag}}\xspace}

\newcommand{\CDP}{\ensuremath{\mathrm{C}_\Dp}\xspace}
\newcommand{\CDS}{\ensuremath{\mathrm{C}_\Ds}\xspace}

\def\acpkkDmethodValue  {13.6} 
\def\acpkkDmethodStat   {8.8}   
\def\acpkkDmethodSyst   {1.6}  

\def\acpkkDsmethodValue  {2.8}
\def\acpkkDsmethodStat   {6.7}   
\def\acpkkDsmethodSyst   {2.0}  

\def\acpkkccorStat {0.05}
\def\acpkkccorSyst {0.28}
\def\acpkkccor {0.06}

\def\acpkkCombinedValue {6.8} 
\def\acpkkCombinedStat {5.4}   
\def\acpkkCombinedSyst {1.6}

\def\adkk  {7.7} 
\def\adkkErr   {5.7}   
\def\adpipi  {23.2} 
\def\adpipiErr   {6.1}   
\def\adhhcor {0.88}

\usepackage{longtable} 
\usepackage{booktabs}
\usepackage{multirow}
\begin{document}

\renewcommand{\thefootnote}{\fnsymbol{footnote}}
\setcounter{footnote}{1}


\begin{titlepage}
\pagenumbering{roman}

\vspace*{-1.5cm}
\centerline{\large EUROPEAN ORGANIZATION FOR NUCLEAR RESEARCH (CERN)}
\vspace*{1.5cm}
\noindent
\begin{tabular*}{\linewidth}{lc@{\extracolsep{\fill}}r@{\extracolsep{0pt}}}
\ifthenelse{\boolean{pdflatex}}
{\vspace*{-1.5cm}\mbox{\!\!\!\includegraphics[width=.14\textwidth]{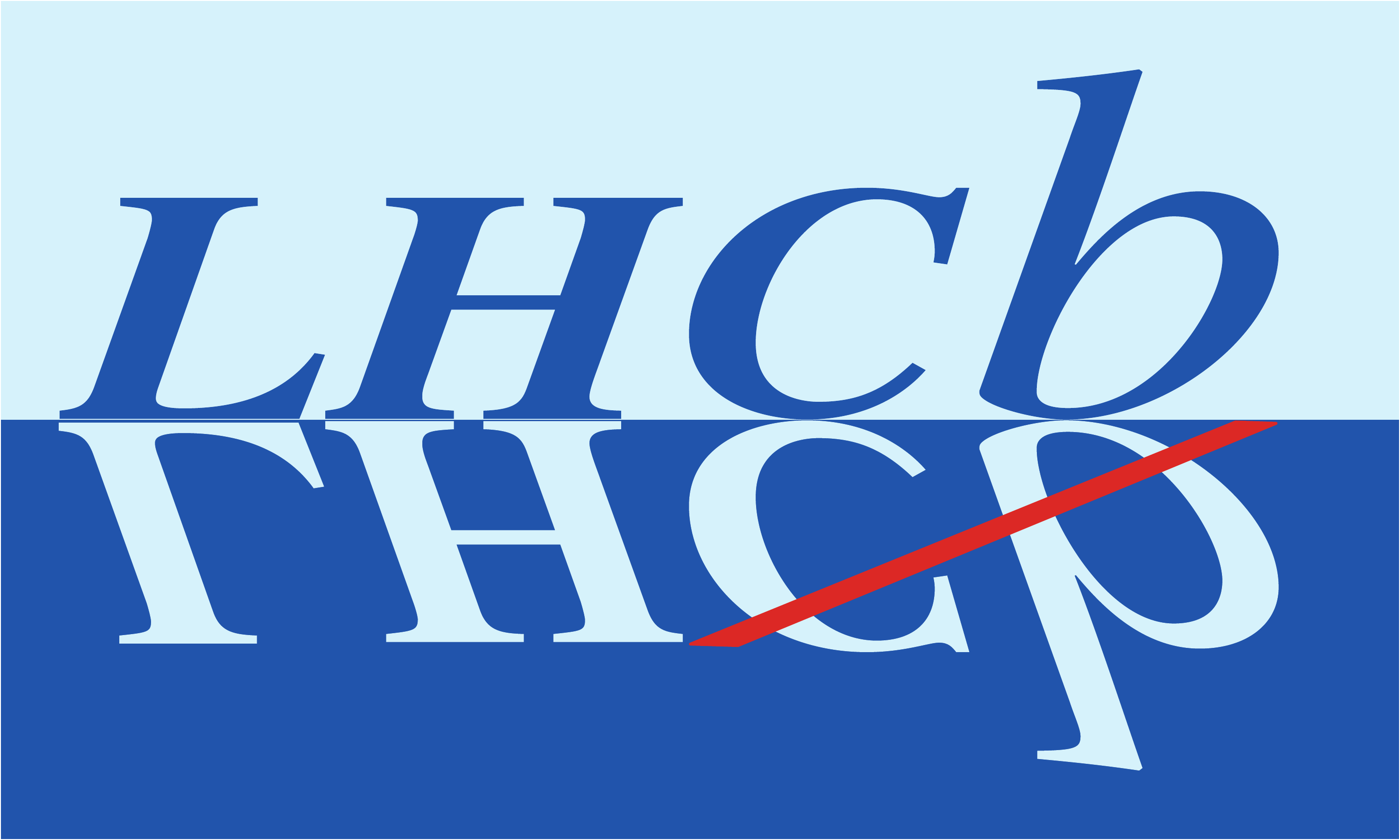}} & &}%
{\vspace*{-1.2cm}\mbox{\!\!\!\includegraphics[width=.12\textwidth]{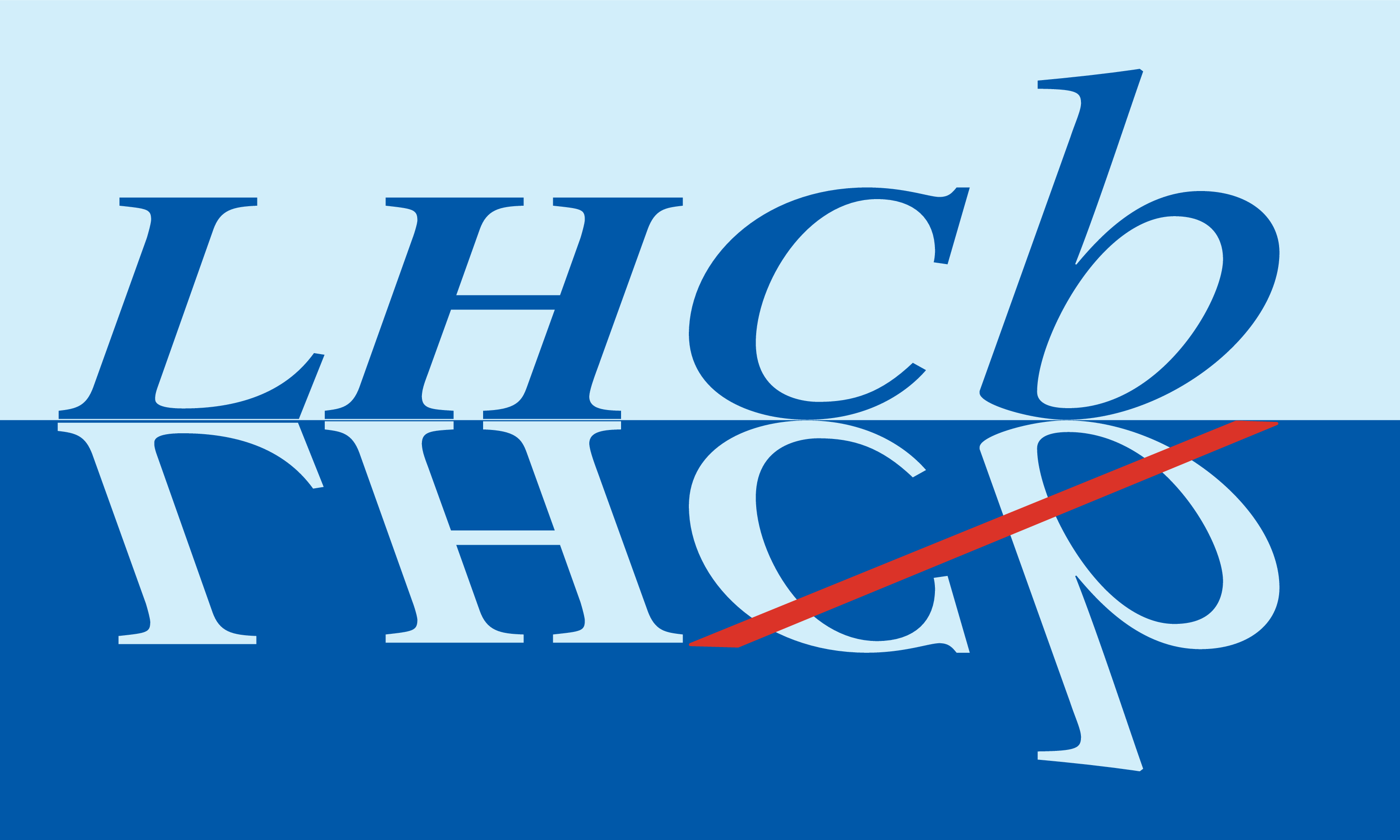}} & &}%
\\
 & & CERN-EP-2022-163 \\  
 & & LHCb-PAPER-2022-024 \\  
 & & September 7, 2022 \\
 & & \\
\end{tabular*}

\vspace*{1.5cm}

{\normalfont\bfseries\boldmath\huge
\begin{center}
  \papertitle 
\end{center}
}

\vspace*{1.5cm}

\begin{center}
\paperauthors
\end{center}

\vspace{\fill}

\begin{abstract}
  \noindent
  The time-integrated \CP asymmetry in the Cabibbo-suppressed decay $\Dz\to\Km\Kp$ is measured using proton-proton collision data, corresponding to an integrated luminosity of $5.7\invfb$ collected at a center-of-mass energy of 13 TeV with the LHCb detector. The \Dz mesons are required to originate from promptly produced $\Dstarp\to\Dz\pip$ decays and the charge of the companion pion is used to determine the flavor of the charm meson at production. 
The time-integrated $\CP$ asymmetry is measured to be
\begin{align*}
    \ACPKK = \left[\acpkkCombinedValue \pm \acpkkCombinedStat \stat \pm \acpkkCombinedSyst \syst\right]\times 10^{-4}.
\end{align*}
The direct $\CP$ asymmetries in $\Dz\to\Km\Kp$ and $\Dz\to\pim\pip$ decays, $a_{\Km\Kp}^{d}$ and  $a_{\pim\pip}^{d}$, are derived by combining $\ACPKK$ with the time-integrated \CP asymmetry difference,
\mbox{$\DACP=\ACPKK-\ACPpipi$}, giving 
\begin{alignat*}{2}
    a_{\Km\Kp}^{d} &= (\phantom{2}\adkk \pm \adkkErr) \times 10^{-4}, \\
    a_{\pim\pip}^{d} &= (\adpipi \pm \adpipiErr) \times 10^{-4},
\end{alignat*}
with a correlation coefficient corresponding to $\rho=\adhhcor$. The compatibility of these results with \CP symmetry is 1.4 and 3.8 standard deviations for $\Dz\to\Km\Kp$ and $\Dz\to\pim\pip$ decays, respectively. This is the first evidence for direct $\CP$ violation in a specific \Dz decay.

\end{abstract}

\vspace*{1.2cm}

\begin{center}
  Published in Phys. Rev. Lett. 131 (2023) 091802
\end{center}

\vspace{0.5cm}

{\footnotesize 
\centerline{\copyright~\papercopyright. \href{\paperlicenceurl}{\paperlicence}.}}
\vspace*{2mm}

\end{titlepage}


\newpage
\setcounter{page}{2}
\mbox{~}
%
%
%
%

\renewcommand{\thefootnote}{\arabic{footnote}}
\setcounter{footnote}{0}
\cleardoublepage
\pagestyle{plain} 
\setcounter{page}{1}
\pagenumbering{arabic}
\noindent One of the three necessary conditions for baryon asymmetry in the Universe is the noninvariance of the fundamental interactions under the simultaneous transformation of the charge conjugation ($C$) and parity ($P$) operators, referred to as \CP violation~\cite{Sakharov:1967dj}. The Cabibbo-Kobayashi-Maskawa (CKM) formalism describes \CP violation in the Standard Model (SM) of particle physics~\cite{Cabibbo:1963yz,Kobayashi:1973fv} through an irreducible phase in the quark-mixing matrix. Over the past sixty years, \CP violation has been observed in the $K$, $D$, and $B$-meson systems by several experiments~\cite{Christenson:1964fg,AlaviHarati:1999xp,Lai:2001ki,Aubert:2001nu,Abe:2001xe,Aubert:2004qm,Chao:2004mn,LHCb-PAPER-2013-018,LHCb-PAPER-2012-001,LHCb-PAPER-2019-006}. In the charm quark sector, the recent observation of \CP violation~\cite{LHCb-PAPER-2019-006} stimulates a wide discussion to understand its nature.
Further precise measurements may resolve the intricate theoretical debate on whether the observed value is consistent with the SM~\cite{Grossman:2006jg,Li:2012cfa,Cheng:2012wr,Khodjamirian:2017zdu,Chala:2019fdb,Grossman:2019xcj,Buccella:2019kpn,Li:2019hho,Schacht:2021jaz,Cheng:2019ggx,Dery:2019ysp,Wang:2020gmn,Bause:2020obd,Dery:2021mll,Cheng:2021yrn,Bediaga:2022sxw}.
The discovery measurement of \CP violation in neutral charm meson decays used the difference between two time-integrated \CP-violating asymmetries of Cabibbo-suppressed $\Dz$ decays, \mbox{$\DACP = \ACP(\KK) - \ACP(\pipi)$}, found to be $\DACP = (-15.4 \pm 2.9)\times 10^{-4}$~\cite{LHCb-PAPER-2019-006}. The time-integrated $\CP$ asymmetry for $f=\KK$ and $f=\pipi$ corresponds to
\begin{align}\label{eq:acp_integrated}
\Acp(f)\ifthenelse{\boolean{prl}}{\!}{}\equiv\ifthenelse{\boolean{prl}}{\!}{}\frac{\int{\deriv t
    \,\epsilon(t)\left[\Gamma(\Dz \to f)(t)\ifthenelse{\boolean{prl}}{\!}{}-\ifthenelse{\boolean{prl}}{\!}{}\Gamma(\Dzb \to
    f)(t)\right]}}{\int{\deriv t\, \epsilon(t) \left[\Gamma(\Dz \to f)(t)\ifthenelse{\boolean{prl}}{\!}{}+\ifthenelse{\boolean{prl}}{\!}{}\Gamma(\Dzb \to f)(t)\right]}},
\end{align}
where $\epsilon(t)$ is the reconstruction efficiency as a function of the \Dz decay time and $\Gamma$ denotes the decay rate.
This Letter presents measurements of the time-integrated \CP asymmetries in $\Dz\to\Km\Kp$ decays. Combining the measurements of $\ACPKK$ and $\DACP$, it is possible to quantify the amount of \CP violation in the decay amplitude for $\Dz\to\Km\Kp$ and $\Dz\to\pim\pip$ decays and provide important insight in the breaking of  $U$-spin symmetry. The mixing in the neutral charm system implies that $\Acp(f)$ is the sum of a component related to the \CP violation in the decay amplitude, $a^d_f$, and a component related to $\Dz$--$\Dzb$ mixing and the interference between mixing and decay, $\Delta Y_f$. Up to first order in the \Dz mixing parameters\cite{BaBar:2010nhz,BaBar:2012bho,Belle:2014ydf,Belle:2015etc,LHCb-PAPER-2017-046,LHCb-PAPER-2018-038,LHCb-PAPER-2019-001,LHCb-PAPER-2021-009}, the time-integrated \CP asymmetry can be written as 
\begin{align}\label{eq:adir}
\Acp(f) \approx a^d_f + \frac{\langle t\rangle_f}{\tau_D} \cdot \Delta Y_f,
\end{align}
where $\langle t\rangle_f$ is the  mean decay time of the $\Dz$ mesons in the experimental data sample and $\tau_D$ is the \Dz lifetime~\cite{Kagan:2020vri,LHCB-PAPER-2020-045}.

The neutral charm mesons considered are produced in the strong-interaction decays $\Dstarp\to\Dz\pip$ from \Dstarp mesons created in proton-proton ($pp$) interactions. The charge of the accompanying ``tagging'' pion (\pitag) is used to identify the flavor of the $\Dz$ meson at production. Throughout this Letter, the inclusion of charge conjugation decay modes is implied, except in the definition of the asymmetries, and \Dstarp and $\phi$ indicate the $\Dstar(2010)^+$ and $\phi(1020)$ mesons, respectively.
The measured asymmetry, $\Ameas(\KK)$, is defined as
\begin{align}
    \label{eq:araw_f_def}
    \Ameas(\KK)\ifthenelse{\boolean{prl}}{\!}{}\equiv\ifthenelse{\boolean{prl}}{\!}{} \frac{N\ifthenelse{\boolean{prl}}{\!}{}\left(\Dstarp\ifthenelse{\boolean{prl}}{\!}{}\to\ifthenelse{\boolean{prl}}{\!}{}\Dz\pip\right) \ifthenelse{\boolean{prl}}{\!}{}-\ifthenelse{\boolean{prl}}{\!}{} N\ifthenelse{\boolean{prl}}{\!}{}\left(\Dstarm\ifthenelse{\boolean{prl}}{\!}{}\to\ifthenelse{\boolean{prl}}{\!}{}\Dzb\pim\right) }{N\ifthenelse{\boolean{prl}}{\!}{}\left(\Dstarp\ifthenelse{\boolean{prl}}{\!}{}\to\ifthenelse{\boolean{prl}}{\!}{}\Dz\pip\right)\ifthenelse{\boolean{prl}}{\!}{}+\ifthenelse{\boolean{prl}}{\!}{} N\ifthenelse{\boolean{prl}}{\!}{}\left(\Dstarm\ifthenelse{\boolean{prl}}{\!}{}\to\ifthenelse{\boolean{prl}}{\!}{}\Dzb\pim\right)},
\end{align}
where $N$ denotes the observed signal yield in the data, and the \Dz meson decays into $\KK$. This asymmetry can be approximated as 
\begin{align}
    \label{eq:araw_f}
    \Ameas(\KK)\ifthenelse{\boolean{prl}}{\!}{} \approx\ifthenelse{\boolean{prl}}{\!}{} \ACP(\KK)+\Aprod(\Dstarp)+\Adet(\pitag),
\end{align}
where $\Aprod(\Dstarp)$ is the production asymmetry arising from the different hadronization probabilities between $\Dstarp$ and $\Dstarm$ mesons in $pp$ collisions, and $\Adet(\pitag)$ is the instrumental asymmetry due to different reconstruction efficiencies of positive and negative tagging pions. The contributions from the production and instrumental asymmetries,  referred to as nuisance asymmetries, are estimated and removed through two calibration procedures denoted as C$_\Dp$ and C$_\Ds$, using a set of promptly produced $\Dp$ and $\Dsp$ meson decays. Namely, the C$_\Dp$ procedure uses $\Dstarp\to\Dz(\to\Km\pip)\pip$, $\Dp\to\Km\pip\pip$ and $\Dp\to\Kzb\pip$ decays; while the C$_\Ds$ procedure uses $\Dstarp\to\Dz(\to\Km\pip)\pip$, $\Dsp\to\phi(\to\KK)\pip$ and $\Dsp\to\Kzb\Kp$ decays. To avoid statistical overlap, the sample of $\Dz\to\Km\pip$ decays is randomly split in two, and the two halves are used separately for the \CDP and \CDS calibration procedures. All these decays are Cabibbo favored, therefore their \CP asymmetries are assumed to be negligible. In analogy to Eq.~\ref{eq:araw_f}, the corresponding measured asymmetries in the calibration decays are decomposed as
\begin{alignat}{2}
\label{eq:A_raws}
&\Ameas(\Km\pip)\ifthenelse{\boolean{prl}}{\!}{} \approx\ifthenelse{\boolean{prl}}{\!}{} \Aprod(\Dstarp)\ifthenelse{\boolean{prl}}{\!}{} -\ifthenelse{\boolean{prl}}{\!}{} \Adet(\Kp) + \Adet(\pip)\ifthenelse{\boolean{prl}}{\!}{} +\ifthenelse{\boolean{prl}}{\!}{} \Adet(\pitag),\nonumber\\ 
&\Ameas(\Km\pip\pip)\ifthenelse{\boolean{prl}}{\!}{} \approx\ifthenelse{\boolean{prl}}{\!}{} \Aprod(\Dp)\ifthenelse{\boolean{prl}}{\!}{} -\ifthenelse{\boolean{prl}}{\!}{} \Adet(\Kp)\ifthenelse{\boolean{prl}}{\!}{} +\ifthenelse{\boolean{prl}}{\!}{} \Adet(\pi^{+}_{1})\ifthenelse{\boolean{prl}}{\!}{} +\ifthenelse{\boolean{prl}}{\!}{} \Adet(\pi^{+}_{2}),\nonumber\\
&\Ameas(\Kzb\pip) \approx \Aprod(\Dp) + A(\Kzb)  + \Adet(\pip),\\
&\Ameas(\phi \pip) \approx \Aprod(\Dsp) + \Adet(\pip), \nonumber\\
&\Ameas(\Kzb \Kp)\approx \Aprod(\Dsp)  + A(\Kzb) + \Adet(\Kp) .\nonumber 
\end{alignat}
In the equations above, $\Adet(\Kp)$ is the kaon instrumental asymmetry, 
$\Aprod(\DporDsp)$ is the $\DporDsp$ meson production asymmetry
and $A(\Kzb)$ is the asymmetry arising from the combined effect of \CP violation and mixing in the neutral kaon system and the different interaction rates of $\Kzb$ and $\Kz$ with the detector material. The asymmetries $\Adet(\pi^{+}_{1})$ and $\Adet(\pi^{+}_{2})$ are related to the two pions in the $\Dp\to\Km\pip\pip$ decay, distinguished by the online selection criteria. In $\Ameas(\phi \pip)$, the asymmetry from the oppositely charged kaons is not included as it is estimated to be negligible. With the individual terms of $\mathcal{O}(10^{-2})$ or less~\cite{LHCb-PAPER-2016-062,LHCb-PAPER-2013-033,LHCb-PAPER-2012-026,LHCb-PAPER-2012-009}, the approximations in Eqs.~\ref{eq:araw_f} and \ref{eq:A_raws} are valid up to corrections of $\mathcal{O}(10^{-6})$. The individual nuisance asymmetries depend on the kinematics of the corresponding particles. After accounting for this kinematic dependence, the time-integrated \CP asymmetry, $\ACPKK$, is obtained for each of the two calibration procedures individually, by combining the measured asymmetries as follows
\begin{alignat}{3}
\label{eq:acpkk}
\ifthenelse{\boolean{prl}}{\CDP\ifthenelse{\boolean{prl}}{\!\!}{}:\,\,& \Acp(\KK) \,&=&\, \Ameas(\KK)-\Ameas(\Km\pip)\nonumber \\ &&+&\Ameas(\Km\pip\pip)\ifthenelse{\boolean{prl}}{\!}{}-\ifthenelse{\boolean{prl}}{\!}{}\Ameas(\Kzb\pip)\ifthenelse{\boolean{prl}}{\!}{}+\ifthenelse{\boolean{prl}}{\!}{}A(\Kzb), \nonumber}{\CDP:\,\,& \Acp(\KK) \,&=&\, \Ameas(\KK)-\Ameas(\Km\pip) +\Ameas(\Km\pip\pip)-\Ameas(\Kzb\pip)+A(\Kzb), \nonumber} \\
\ifthenelse{\boolean{prl}}{\CDS\ifthenelse{\boolean{prl}}{\!\!}{}:\,\,& \Acp(\KK) \,&=&\,  \Ameas(\KK)-\Ameas(\Km\pip)\nonumber \\ &&+&\Ameas(\phi\pip)\ifthenelse{\boolean{prl}}{\!}{}-\ifthenelse{\boolean{prl}}{\!}{}\Ameas(\Kzb\Kp)\ifthenelse{\boolean{prl}}{\!}{}+\ifthenelse{\boolean{prl}}{\!}{}A(\Kzb)}{\CDS:\,\,& \Acp(\KK) \,&=&\,  \Ameas(\KK)-\Ameas(\Km\pip)+\Ameas(\phi\pip)-\Ameas(\Kzb\Kp)+A(\Kzb)}.
\end{alignat}

The asymmetries are measured in $pp$ collision data, collected with the LHCb detector at a center-of-mass energy of $13\tev$, corresponding to an integrated luminosity of $5.7\invfb$. The \lhcb detector is a single-arm forward spectrometer designed for the study
of particles containing \bquark or \cquark quarks~\cite{LHCb-DP-2008-001,LHCb-DP-2014-002}. A high-precision tracking system with a dipole magnet and vertex detector measures the momentum ($p$) and impact parameter (IP) of charged particles. The IP is defined as the distance of closest approach
between the reconstructed trajectory and a $pp$ interaction vertex~\cite{Kucharczyk:1756296}. The IP is used
to distinguish between particles produced in the primary collisions and those produced in heavy-flavor decays.
Different species of charged hadrons 
are distinguished using particle identification (PID) information from two
ring-imaging Cherenkov detectors, an electromagnetic and a hadronic calorimeter, and a muon detector. 

The online event selection, the trigger, consists of a hardware stage followed by two software stages within which a near real-time alignment and calibration of the detector are performed~\cite{LHCb-PROC-2015-011}.
In the hardware stage, events are selected based on calorimeter and muon detector information and are accepted independently of the charm decay of interest, reducing any related asymmetry to a negligible level.  
The subsequent first stage of the software trigger reconstructs the trajectories using information from the full LHCb tracking system and applies requirements on the transverse momentum (\pt), the IP, and the displacement from any primary vertex (PV) of the charm-meson decay products.
To pass the selection, at least one charged particle or two particles forming a high-quality vertex must fulfill these criteria.
The second stage of the software trigger exploits the full information from the tracking sub-detectors and performs additional steps of the pattern recognition, including the reconstruction of neutral particles and PID.
Further requirements on PID, kinematics and the decay topology are then applied.

The $\Dsp\to\phi\pip$ decays are selected from $\Dsp\to\KK\pip$ decay candidates requiring that the invariant mass of the kaon pair must be within $\pm 5$\mevcc of the $\phi$ mass. Similarly, the $\Kzb$ mesons, produced in $\Dp\to\Kzb\pip$ and $\Dsp\to\Kzb\Kp$ decays, are reconstructed using their decay to two pions, which is dominated by the $\KS$ state. The two pions are required to have an invariant mass within $\pm$10\mevcc of the \KS mass and to form a vertex displaced more than 20\mm along the beam direction from the \DporDsp-meson decay. The $\Dz$ candidates are required to have a reconstructed invariant mass between 1844 and 1887~\mevcc.

An offline selection is applied to reduce background, including combinations of random tracks and tracks from other $\cquark$-hadron decays, and to ensure a further cancellation of nuisance asymmetries which can depend on the kinematics of the charm mesons, the kaons, and the pions. These kinematics and PID requirements are applied to both the signal and related control modes where applicable. To improve the overall precision, these selections have been optimized independently for each of the two calibration sets C$_\Dp$ and C$_\Ds$.
A requirement on the IP of the charm hadron suppresses charm mesons from \bquark-hadron decays to a fraction between 2\% and 6\% in all decay modes.
To improve the resolution on the track momenta and the charm meson decay length and invariant mass, a global decay-chain fit~\cite{Hulsbergen:2005pu} is performed, constraining the origin vertex of the charm meson to the position of the nearest primary vertex and the invariant mass of the two pion system to the known $\KS$ mass~\cite{PDG2022}.

In the construction of the \Dstarp candidate, requirements are imposed on the tagging pion to exclude kinematic regions which show a large asymmetry in $\Adet(\pitag)$~\cite{LHCb-PAPER-2015-055}.
The invariant mass of the \Dstarp, $m(\Dz\pip)$, calculated using the vector sum of the momenta of the three charged particles and the known \Dz and \pip masses~\cite{PDG2022}, is required to be between 2004.5 and 2020\mevcc.
For events that contain multiple \Dstarp candidates, one candidate is retained randomly.

\begin{figure}[]
\centering
\ifthenelse{\boolean{wordcount}}{}{
\ifthenelse{\boolean{prl}}{\includegraphics[width=0.45\textwidth]{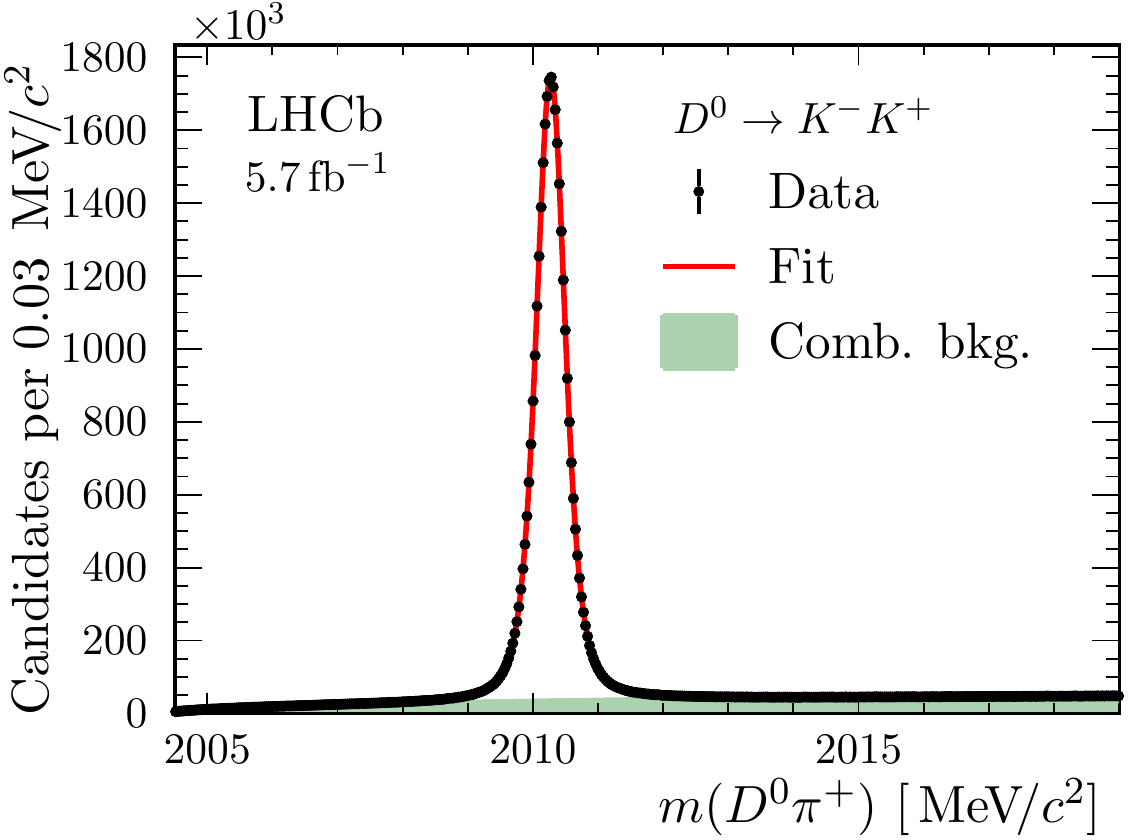}}{\includegraphics[width=0.90\textwidth]{figs/Fig1.pdf}}
}
\caption{Distribution of the invariant mass for the weighted $\mbox{\Dstarp\to\Dz(\to\Km\Kp)\pip}$ decay candidates, from the C$_\Dp$ calibration procedure. The result of the fit to this distribution is also shown. }
\label{fig:selection_mass_distributions}
\end{figure}

The nuisance asymmetries introduced in Eqs.~\ref{eq:araw_f} and \ref{eq:A_raws} are expected to depend on the kinematics of the individual particles. To ensure a proper cancellation of those asymmetries, per-candidate weights are applied to all the data samples to equalize the kinematics  of $\Dstarp$, $\Dp$ and $\Dsp$ mesons and the kaons and pions~\cite{sup}. The values of the weights are calculated separately for each calibration procedure using an iterative technique.
It is verified that the background-subtracted, weighted distributions of the components of momenta of the relevant particles agree among the different decays.
The weighting procedure is repeated for each data-taking year and magnet polarity to account for the dependence of the nuisance asymmetries on data-taking conditions.

The measured asymmetries of signal components for each decay mode are determined through least-square fits to the weighted, binned mass distributions of the charm-meson candidates, simultaneously for both flavors~\cite{sup}.
The signal models consist of a sum of Gaussian and Johnson $S_U$ functions~\cite{Johnson:1949zj}, empirically describing the experimental resolution and the energy loss due to final-state radiation.
The means of the signal distributions are distinct for the two charm meson flavors, whereas all the other parameters, including the relative fractions among the various functions, are shared. For \Dstarp decays, the combinatorial background is described by an empirical function of the form $[m(\Dz\pip) - m(\Dz) - m(\pip)]^\alpha e^{\beta m(\Dz\pip)}$, where $\alpha$ and $\beta$ are two parameters shared between the two flavors. In the other cases, an exponential function with a distinct parameter for positive and negative particles is used.

Figure~\ref{fig:selection_mass_distributions} presents the distribution of the $\Dz\to\Km\Kp$ invariant mass and the result of the fit. The signal yields, together with the statistical reduction factor, defined as $(\Sigma_{i=1}^{i=K} w_i)^2/(N\cdot\Sigma_{i=1}^{i=K} w_i^2)$, where $K$ is the total number of candidates and $w_i$ includes background subtraction and kinematic weights, are reported in Table~\ref{tab:yields}. These reduction factors are for illustrative purposes only and indicate the hypothetical fraction of signal events that would provide the same statistical power as the weighted data sample.

\begin{table}[t]
\setlength{\tabcolsep}{1em}\centering
\caption{Signal yields and statistical reduction factors arising from the kinematic weighting of the sample for the various decay modes and both calibration procedures.}
\ifthenelse{\boolean{wordcount}}{}{
\bgroup
\def\arraystretch{1.2}
\begin{tabular}{p{8em} c c c c}
\toprule 
Decay mode & \multicolumn{2}{c}{Signal yield [$10^6$]} & \multicolumn{2}{c}{Red. factor } \\
\cmidrule(r{.5em}){2-3} \cmidrule(l{.5em}){4-5} 
& C$_\Dp$ & C$_\Ds$  & C$_\Dp$ &  C$_\Ds$   \\
\midrule
$\Dz\to\Km\Kp$       & 37 & 37 & 0.72 & 0.76\\
$\Dz\to\Km\pip$    & 58 & 56 & 0.33 & 0.76\\
$\Dp\to\Km\pip\pip$ & 188 & -- & 0.23 & --\\
$\Dp\to\Kzb\pip$    & 6 & -- & 0.25 & --\\
$\Dsp\to\phi\pip$  & -- & 43 & -- & 0.55\\
$\Dsp\to\Kzb\Kp$ & -- & 5 & -- & 0.70\\
\bottomrule
\end{tabular}
\egroup
}
\label{tab:yields}
\end{table}

Separate fits are performed to subsamples of data collected in different years and with different magnet polarities. After determining the asymmetries in these subsamples, the values of \acpkk are calculated according to Eq.~\ref{eq:acpkk}, taking into account the contribution from the neutral kaon asymmetry. This is estimated by combining the \lhcb material map from simulation with measured \CP-violation and cross-section parameters of the neutral kaon system~\cite{Fetscher:1996fa,Gsponer:1978dt,Briere:1995tw}, following the procedure described in Ref.~\cite{LHCb-PAPER-2014-013}.
The correction considers different \Kzb momentum spectra for the C$_\Dp$ and C$_\Ds$ calibration procedure and corresponds to $(-5.1 \pm 0.6)\times 10^{-4}$ and $(-8.5\pm 1.3)\times 10^{-4}$, respectively. The uncertainties are evaluated with a model-independent strategy based on data and discussed later in the Letter.
The individual \ACPKK values per subsample are found to be in agreement, with a $p$-value of 0.85 and 0.22 for the C$_\Dp$ and C$_\Ds$ methods, respectively. Finally, the measurements in each subsample are averaged to obtain the final result for each procedure. 

Several sources of systematic uncertainties are considered.
The systematic uncertainty related to the description of signal and background in the invariant-mass distributions is evaluated by generating pseudoexperiments according to the baseline fit models, and fitting alternative models to those samples. A fit-independent approach is also considered, based on a sideband subtraction. 
Systematic uncertainties  of $1.1 \times 10^{-4}$  and $1.0 \times 10^{-4}$ are assigned for the C$_\Dp$ and C$_\Ds$ procedures, with a correlation of 0.05.

A systematic uncertainty associated to the presence of background components peaking in $m(\Dz \pi)$ and not in $m(\KK)$ is determined by fitting the latter distribution in the $\Dz \to \KK$ samples~\cite{sup}.
Various backgrounds are modeled using fast simulation~\cite{Cowan:2016tnm}.
The main sources are $\Dz \to \Km \pip \piz$ and $\Dz \to \Km \ep \nu_e$ decays.
A similar study is performed on the $\Dz \to \Km\pip$ decay sample, where the peaking-background contributions are found to be negligible.
As a result, the values $0.3 \times 10^{-4}$ and $0.4 \times 10^{-4}$ are assigned as systematic uncertainties for the C$_\Dp$ and C$_\Ds$ calibration procedures, respectively, with a correlation coefficient of 0.74.

Although suppressed by the stringent requirement on the IP, a fraction of \D mesons from $\bquark$-hadron decays is still present in the final sample.
As the different decay modes may have different levels of contamination, the value of $\ACPKK$ may be affected by an incomplete cancellation of the production asymmetries of $\bquark$-hadrons.
The contributions from $b$-hadron decays in data are estimated by fitting the IP distribution of charm mesons using shapes obtained from simulation.
The corresponding systematic uncertainties are estimated to be $0.6 \times 10^{-4}$ and $0.3 \times 10^{-4}$ for the C$_\Dp$ and C$_\Ds$ calibration procedures, respectively, with a negligible correlation between them.

Any residual disagreement between the kinematic distributions among the various decay modes leads to an imperfect cancellation of the nuisance asymmetries.
The systematic uncertainties related to this effect are estimated to be $0.8 \times 10^{-4}$ and $0.4 \times 10^{-4}$ for the C$_\Dp$ and C$_\Ds$ procedures, respectively, with a negligible correlation.

To test the accuracy of the estimated value for $A(\Kzb)$, a linear term with one free parameter is introduced in the model that describes the dependence of $A(\Kzb)$ on the neutral-kaon decay time~\cite{sup}. The parameter is determined by fitting the charge asymmetry in $\Dp\to\Kzb\pip$ decays as a function of the \Kzb decay time. This is done using a control sample where the neutral kaon decays outside the vertex detector.
The parameter is found to be consistent with zero. Its uncertainty is propagated to the $\Kzb$ lifetimes relevant for $\ACPKK$ and assigned as systematic uncertainty.
The resulting, fully correlated, systematic uncertainties are $0.6 \times 10^{-4}$ and $1.3 \times 10^{-4}$ for the C$_\Dp$ and C$_\Ds$ procedures, respectively.

In the C$_\Ds$ procedure, $\Dsp \to \KK \pip$ decay modes other than $\Dsp \to \phi \pip$ may break the symmetry between the $\Km$ and $\Kp$ meson kinematic distributions.
This leads to a bias in the measured asymmetry due to the momentum-dependent instrumental asymmetry of the kaon.
This effect is estimated by combining the two momentum distributions with the expected charged-kaon asymmetry from simulation. The resulting systematic uncertainty is $1.0 \times 10^{-4}$.

All individual contributions are summed in quadrature to give the total systematic uncertainties of $1.6 \times 10^{-4}$ and $2.0 \times 10^{-4}$ for the C$_\Dp$ and C$_\Ds$ procedures, respectively.
A summary of all systematic uncertainties is shown in Table~\ref{tab:syst_summary}.

\begin{table}[t]
\centering %
\caption{
    Systematic uncertainties on \acpkk for the two calibration procedures C$_\Dp$ and C$_\Ds$. The total uncertainties are obtained as the sums in quadrature of the individual contributions. Correlations between the systematic uncertainties of the two calibration procedures are also reported.}
\ifthenelse{\boolean{wordcount}}{}{
\bgroup
\def\arraystretch{1.2}
\begin{tabular}{l c c c}
\toprule 
Source & C$_\Dp$ [$10^{-4}$] & C$_\Ds$ [$10^{-4}$] & Corr. \\
\midrule
Fit model              & 1.1 & 1.0 & 0.05\\
Peaking backgrounds    & 0.3 & 0.4 & 0.74\\
Secondary decays       & 0.6 & 0.3 & --\\
Kinematic weighting    & 0.8 & 0.4 & --\\
Neutral kaon asymmetry & 0.6 & 1.3 & 1.00\\
Charged kaon asymmetry & --  & 1.0 & --\\
\midrule
Total                  & 1.6 & 2.0 & 0.28 \\
\bottomrule
\end{tabular}
\egroup
}
\label{tab:syst_summary}
\end{table}

Numerous additional checks are carried out~\cite{sup}.
The measurements of \ACPKK are verified to not depend on the decay time, the transverse momentum and the pseudorapidity of the \Dz meson; the decay time and the pseudorapidity of the \Kzb meson; and the IP significance of the final-state particles with respect to all the PVs in the event of the control modes. The IP significance is defined  as the difference between the $\chisq$ of the PV reconstructed with and without the considered particle.
Furthermore, the total sample is split by different data-taking periods, also distinguishing different magnet polarities.
Splitting into subsamples based on the trigger configuration is also considered.
The $p$-values under the hypothesis of no dependencies of \ACPKK on the various  variables are found to be uniformly distributed.
Checks using alternative PID requirements and trigger selections are performed, and all variations of \ACPKK are found to be compatible within statistical uncertainties.
The resulting values for $\ACPKK$ for both calibration procedures are
\begin{alignat}{4} 
\text{C$_\Dp$}\ifthenelse{\boolean{prl}}{\!\!}{}&:\ifthenelse{\boolean{prl}}{}{\,\,} \acpkk  \,&=&\, \left[ \acpkkDmethodValue\ifthenelse{\boolean{prl}}{\!}{}\pm\ifthenelse{\boolean{prl}}{\!}{} \acpkkDmethodStat\ifthenelse{\boolean{prl}}{\!}{}\stat\ifthenelse{\boolean{prl}}{\!}{}\pm\ifthenelse{\boolean{prl}}{\!}{} \acpkkDmethodSyst \ifthenelse{\boolean{prl}}{\!}{}\syst\right]\ifthenelse{\boolean{prl}}{\!}{}\times\ifthenelse{\boolean{prl}}{\!}{} 10^{-4},\nonumber\\
\text{C$_\Ds$}\ifthenelse{\boolean{prl}}{\!\!}{}&:\ifthenelse{\boolean{prl}}{}{\,\,} \acpkk &=&\, \left[ \phantom{0}\acpkkDsmethodValue\ifthenelse{\boolean{prl}}{\!}{}\pm\ifthenelse{\boolean{prl}}{\!}{}\acpkkDsmethodStat \ifthenelse{\boolean{prl}}{\!}{}\stat\ifthenelse{\boolean{prl}}{\!}{}\pm\ifthenelse{\boolean{prl}}{\!}{} \acpkkDsmethodSyst\ifthenelse{\boolean{prl}}{\!}{}\syst \right]\ifthenelse{\boolean{prl}}{\!}{}\times\ifthenelse{\boolean{prl}}{\!}{} 10^{-4}\nonumber,
\end{alignat}
with a statistical and systematic correlations of $\acpkkccorStat$
and $\acpkkccorSyst$ respectively, corresponding to a total correlation of $\acpkkccor$. 
The two results are in agreement within one standard deviation. Their average is
\begin{align}
\ACPKK = \left[\acpkkCombinedValue \pm \acpkkCombinedStat \stat  \pm \acpkkCombinedSyst \syst \right]\times 10^{-4},\nonumber
\end{align}
consistent with the previous results~\cite{LHCb-PAPER-2016-035,LHCb-PAPER-2014-013,sup}.
Assuming that \CP is conserved in mixing and in the interference between decay and mixing, the comparison of the result reported here with the current world average~\cite{HFLAV18} gives a compatibility of 1.3 standard deviations. 

A combination of all the time-integrated \CP asymmetries measured by the LHCb collaboration to date is performed, under the hypothesis that the time-dependent \CP violation term in Eq.~\ref{eq:adir} is final-state independent, \ie $\Delta Y_{\KK}=\Delta Y_{\pipi}=\Delta Y$, as the final-state dependent contributions are estimated to be of the order of~$10^{-5}$~\cite{LHCB-PAPER-2020-045}. The combination includes the previous LHCb measurements of \ACPKK~\cite{LHCb-PAPER-2016-035,LHCb-PAPER-2014-013} and \DACP~\cite{LHCb-PAPER-2019-006,LHCb-PAPER-2015-055,LHCb-PAPER-2014-013} as well as the current LHCb average of $\Delta Y$~\cite{LHCB-PAPER-2020-045}, the world average of the \Dz lifetime~\cite{PDG2022} 
and the values of reconstructed mean decay times for the $\Dz\to\KK$ and $\Dz\to\pipi$ decays in the various analysis. 
The combination, obtained by minimizing a $\chisq$ function that includes all the measurements and their correlations, leads to 
\begin{alignat*}{2}
    a_{\KK}^{d} &= (\phantom{0}\adkk \pm \adkkErr) \times 10^{-4}, \\
    a_{\pipi}^{d} &= (\adpipi \pm \adpipiErr) \times 10^{-4},
\end{alignat*}
where the uncertainties include systematic and statistical contributions with a correlation coefficient of $\adhhcor$. 
\begin{figure}[ht!]
\centering
\ifthenelse{\boolean{wordcount}}{}{
\ifthenelse{\boolean{prl}}{\includegraphics[width=0.45\textwidth]{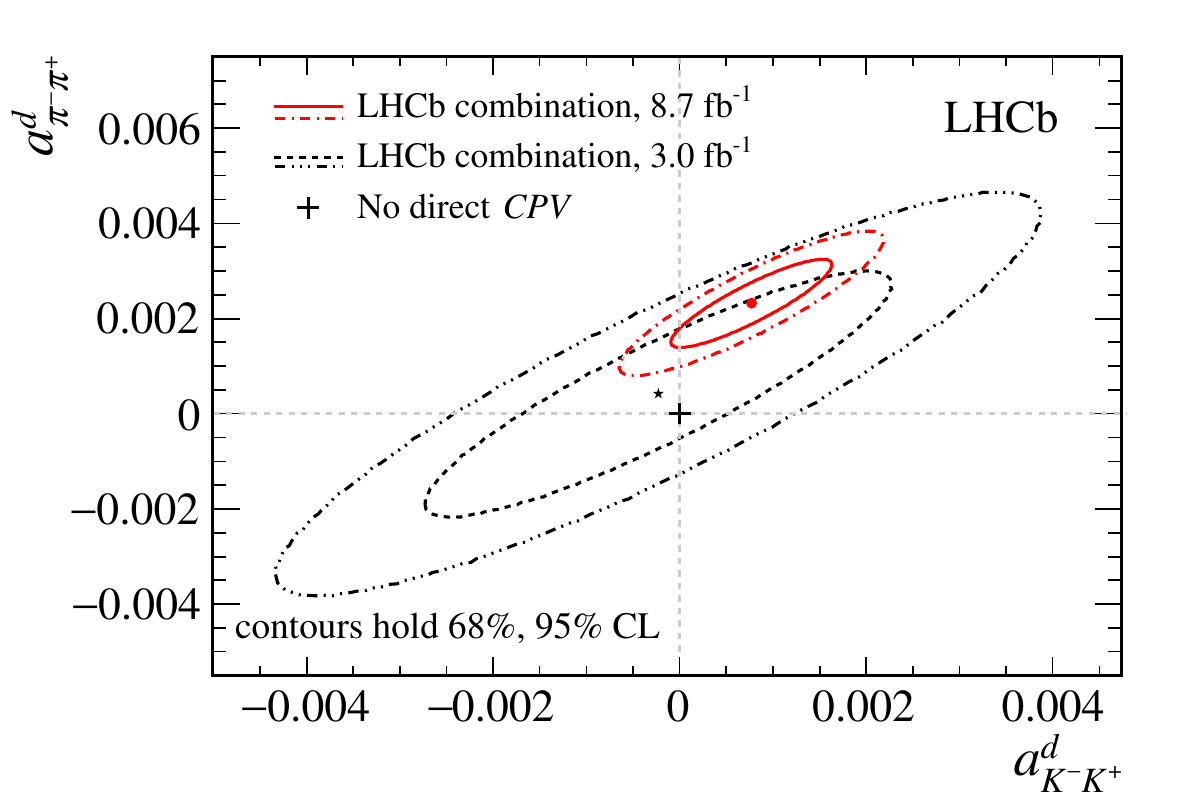}\hfil}{\includegraphics[width=0.90\textwidth]{figs/Fig2.pdf}\hfil} 
}
\caption{Central values and two-dimensional confidence regions in the $(a_{\KK}^{d},a_{\pipi}^{d})$ plane for the combinations of the LHCb results obtained with the dataset taken between 2010 and 2018 and the one taken between 2010 and 2012, corresponding to an integrated luminosity of 8.7\invfb and 3.0\invfb, respectively.}\label{fig:countour}
\end{figure}
Figure~\ref{fig:countour} shows the central values and the confidence regions in the $(a_{\KK}^{d},a_{\pipi}^{d})$ plane for this combination and the one realized with data collected between 2010 and 2012~\cite{LHCb-PAPER-2015-055,LHCb-PAPER-2016-035,LHCb-PAPER-2014-013,LHCb-PAPER-2014-069,LHCb-PAPER-2016-063}. The two combinations are based on an integrated luminosity of 8.7 fb$^{-1}$ and 3.0 fb$^{-1}$, respectively. 

The direct \CP asymmetries deviate from zero by 1.4 and 3.8 standard deviations for $\Dz\to\KK$ and $\Dz\to\pipi$ decays, respectively. This is the first evidence for direct \CP violation in the $\Dz\to\pipi$ decay. $U$-spin symmetry implies $a^d_{\KK} + a^d_{\pipi}=0$~\cite{Gronau:2013xba}. A value of \mbox{$a^d_{\KK} + a^d_{\pipi} = (30.8 \pm 11.4 ) \times 10^{-4}$} has been found, corresponding to a departure from $U$-spin symmetry of 2.7 standard deviations.

In summary, this Letter reports the most precise measurement of the time-integrated \CP asymmetry in the $\Dz\to\KK$ decay to date. A combination with the previous LHCb measurements shows the first evidence of direct \CP asymmetry in an individual charm meson decay. These results will help to clarify the theoretical understanding of whether the observed \CP violation in neutral charm meson decays is consistent with the SM, or an indication of the existence of new dynamics.

\section*{Acknowledgements}
%
%
\noindent We express our gratitude to our colleagues in the CERN
accelerator departments for the excellent performance of the LHC. We
thank the technical and administrative staff at the LHCb
institutes.
We acknowledge support from CERN and from the national agencies:
CAPES, CNPq, FAPERJ and FINEP (Brazil); 
MOST and NSFC (China); 
CNRS/IN2P3 (France); 
BMBF, DFG and MPG (Germany); 
INFN (Italy); 
NWO (Netherlands); 
MNiSW and NCN (Poland); 
MEN/IFA (Romania); 
MICINN (Spain); 
SNSF and SER (Switzerland); 
NASU (Ukraine); 
STFC (United Kingdom); 
DOE NP and NSF (USA).
We acknowledge the computing resources that are provided by CERN, IN2P3
(France), KIT and DESY (Germany), INFN (Italy), SURF (Netherlands),
PIC (Spain), GridPP (United Kingdom), 
CSCS (Switzerland), IFIN-HH (Romania), CBPF (Brazil),
Polish WLCG  (Poland) and NERSC (USA).
We are indebted to the communities behind the multiple open-source
software packages on which we depend.
Individual groups or members have received support from
ARC and ARDC (Australia);
Minciencias (Colombia);
AvH Foundation (Germany);
EPLANET, Marie Sk\l{}odowska-Curie Actions and ERC (European Union);
A*MIDEX, ANR, IPhU and Labex P2IO, and R\'{e}gion Auvergne-Rh\^{o}ne-Alpes (France);
Key Research Program of Frontier Sciences of CAS, CAS PIFI, CAS CCEPP, 
Fundamental Research Funds for the Central Universities, 
and Sci. \& Tech. Program of Guangzhou (China);
GVA, XuntaGal, GENCAT and Prog.~Atracci\'on Talento, CM (Spain);
SRC (Sweden);
the Leverhulme Trust, the Royal Society
 and UKRI (United Kingdom).

\clearpage
\clearpage
\section*{Supplemental material}
\label{sec:supp-PRL}
\subsection*{Reconstructed mean decay times}

The interpretation of \acpkk in terms of direct \CP asymmetries, $a^d_{\pim\pip}$ and $a^d_{\Km\Kp}$ requires the measurement of the reconstructed mean decay time of $\Dz \to\Km\Kp$ decay. The values corresponding to the measurements presented in this Letter are \mbox{$\langle t\rangle_{\Km\Kp}=(7.315 \pm 0.020)\times 10^{-13}$~s} and $\langle t\rangle_{\Km\Kp}=(6.868 \pm 0.014)\times 10^{-13}$~s for the \CDP and \CDS methods, respectively. Their correlation corresponds to $\rho=0.74$. These measurements are also correlated with the difference of reconstructed mean decay times for $\Dz \to\Km\Kp$ and $\Dz \to\pim\pip$ decays, $\Delta\langle t\rangle^{\pi-\mathrm{tagged}}$, measured in Ref.~\cite{LHCb-PAPER-2019-006}. The correlation coefficients between $\langle t\rangle_{\Km\Kp}$ and $\Delta\langle t\rangle^{\pi-\mathrm{tagged}}$  are $\rho=0.23$ and $\rho=0.25$ for the \CDP and \CDS methods, respectively.     

\subsection*{Additional plots}




\begin{figure}[htb!]
\centering
\includegraphics[width=0.49\textwidth]{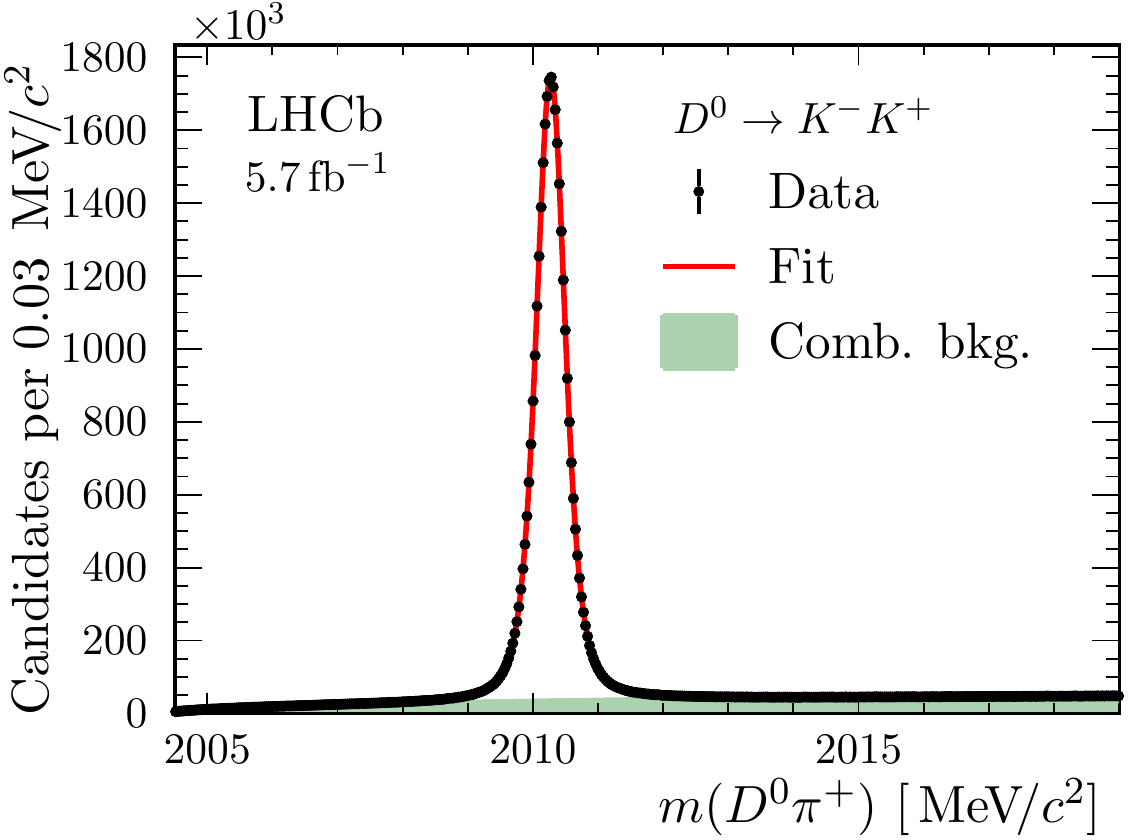}\hfil
\includegraphics[width=0.49\textwidth]{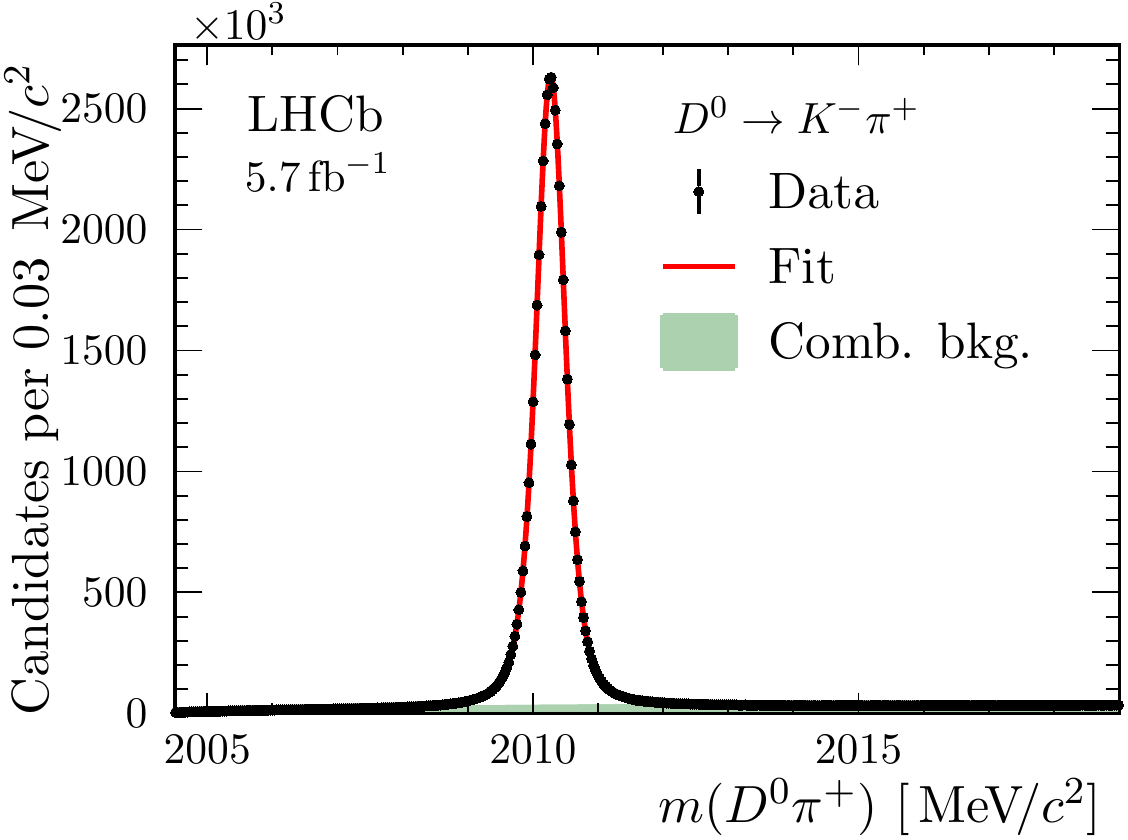}\\
\includegraphics[width=0.49\textwidth]{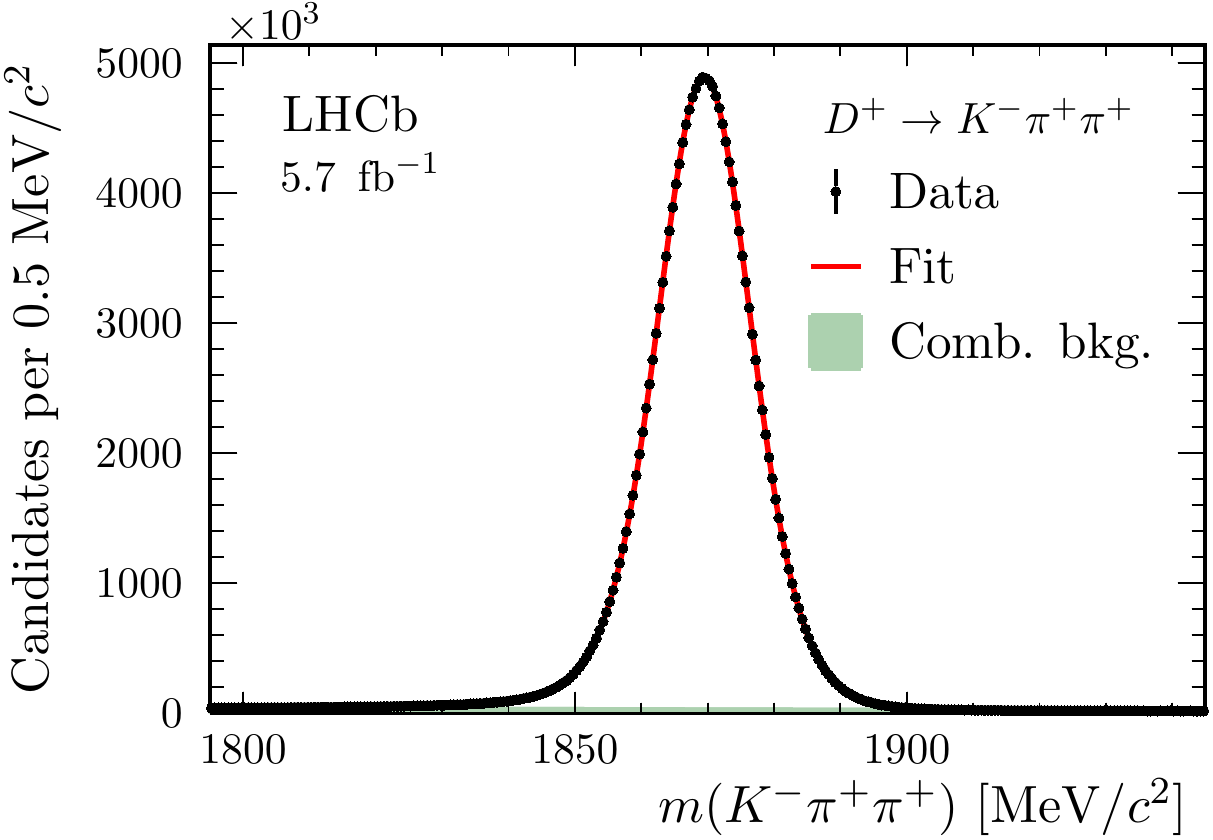}\hfil
\includegraphics[width=0.49\textwidth]{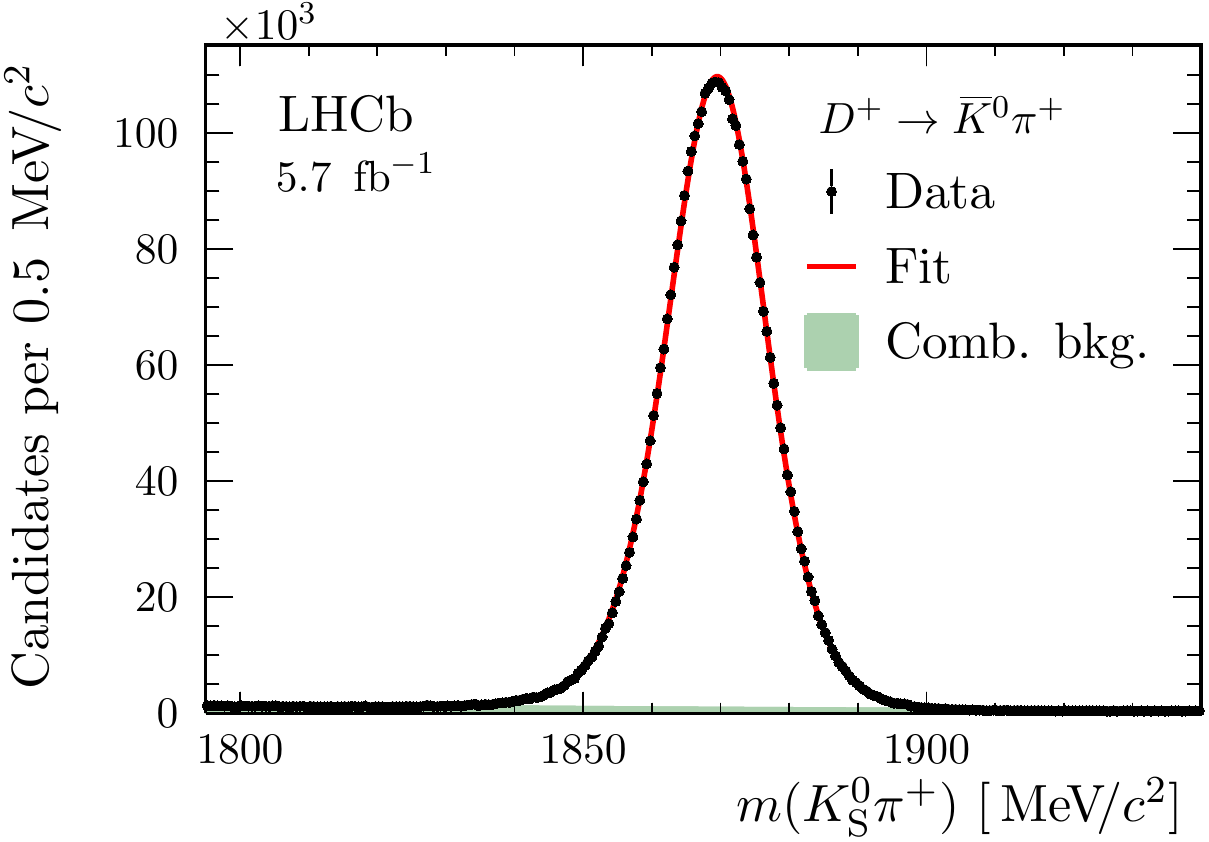}\\
\caption{Distributions of the invariant mass for the weighted charm-meson candidates, for
the decays (top left) \Dstarp\to\Dz(\to\Km\Kp)\pip, (top right) \Dstarp\to\Dz(\to\Km\pip)\pip, (bottom
left) \Dp\to\Km\pip\pip, (bottom right) \Dp\to\Kzb\pip. The data are from the
\CDP calibration procedure. The results of the fits to these
distributions are also shown. The top left is repeated from the main text.}
\end{figure}

\begin{figure}[htb!]
\centering
\includegraphics[width=0.49\textwidth]{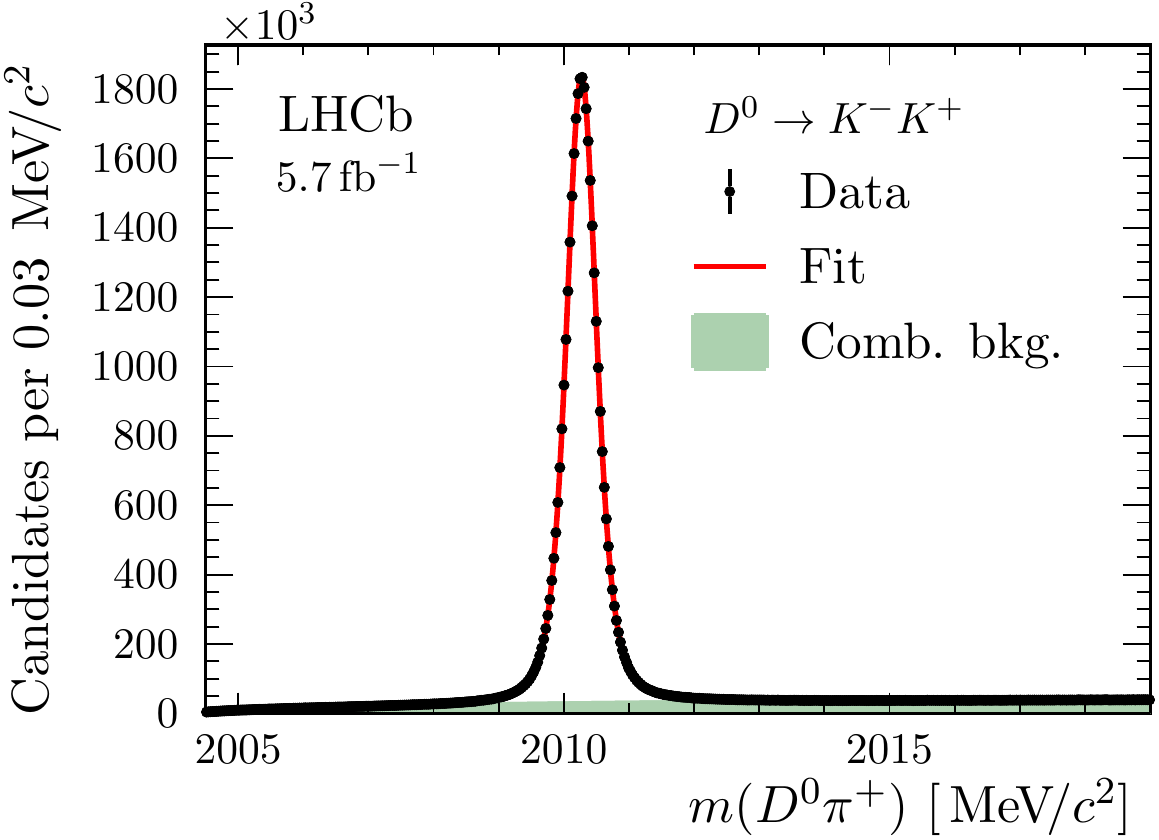}\hfil
\includegraphics[width=0.49\textwidth]{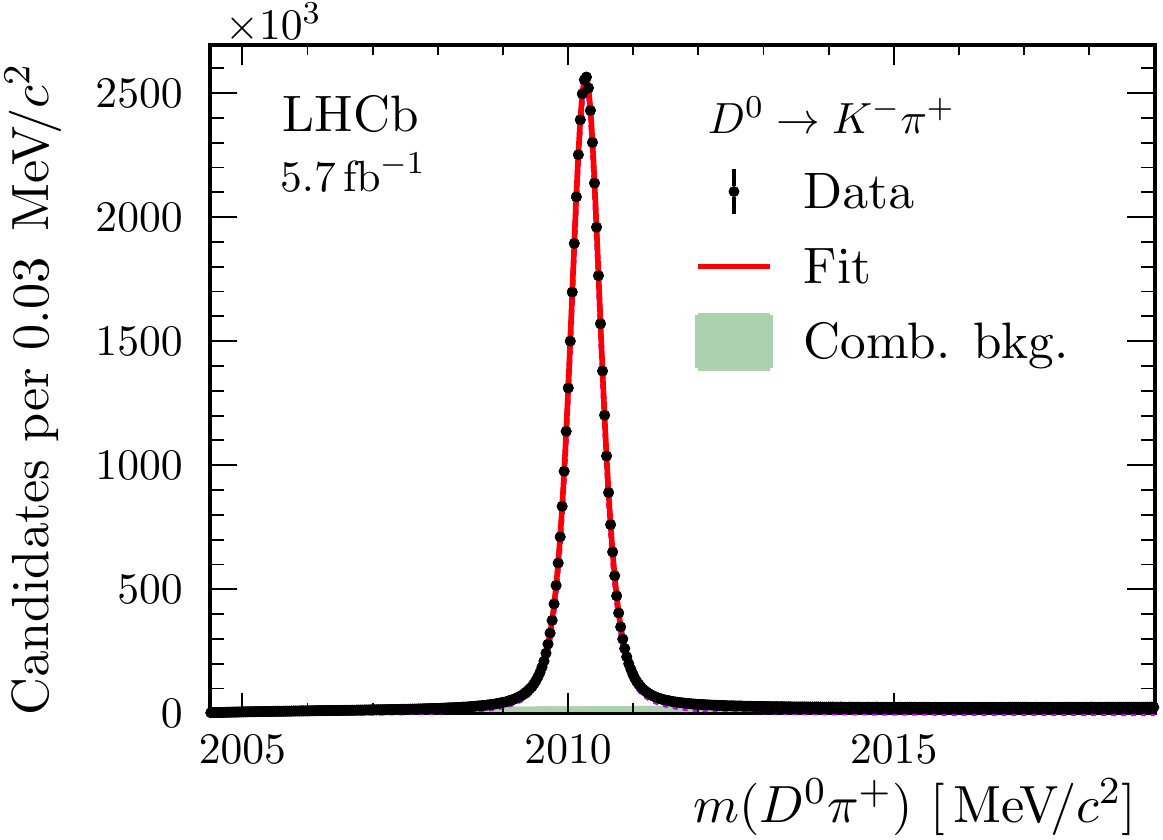}
\includegraphics[width=0.49\textwidth]{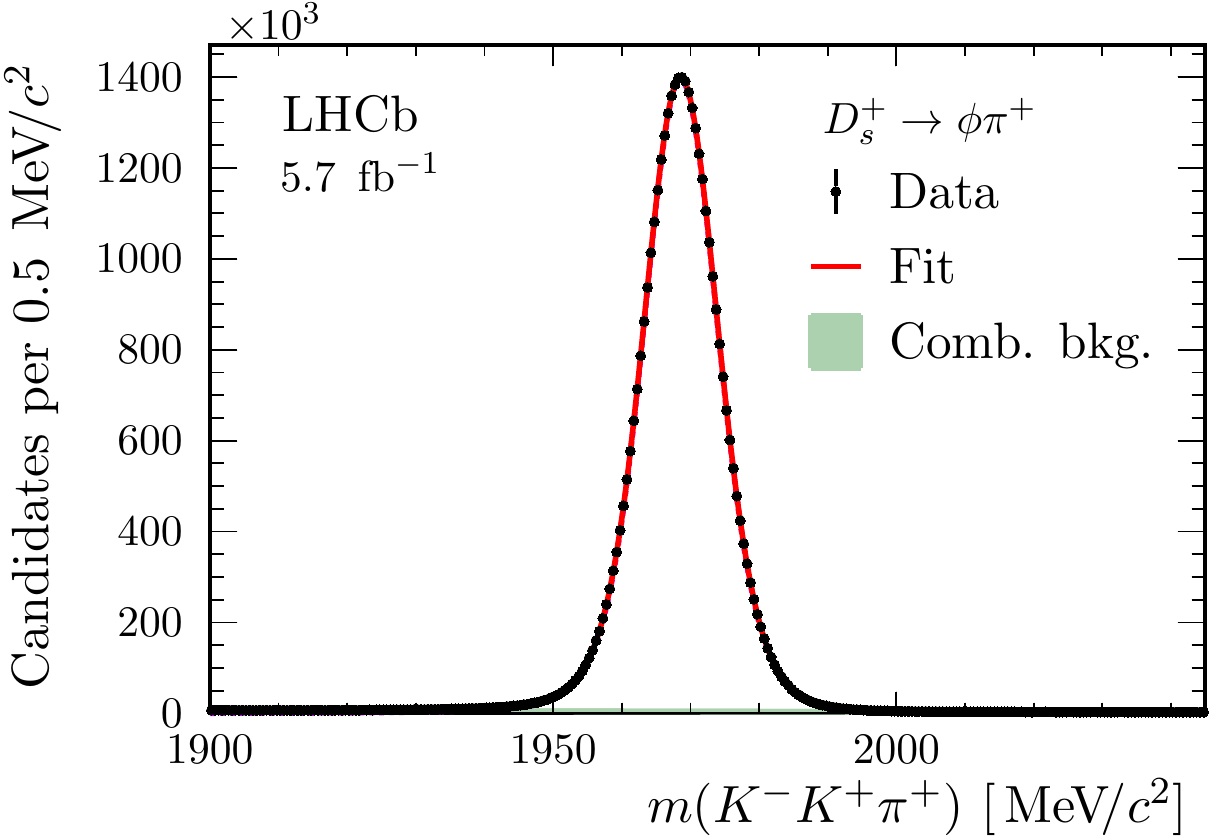}\hfil
\includegraphics[width=0.49\textwidth]{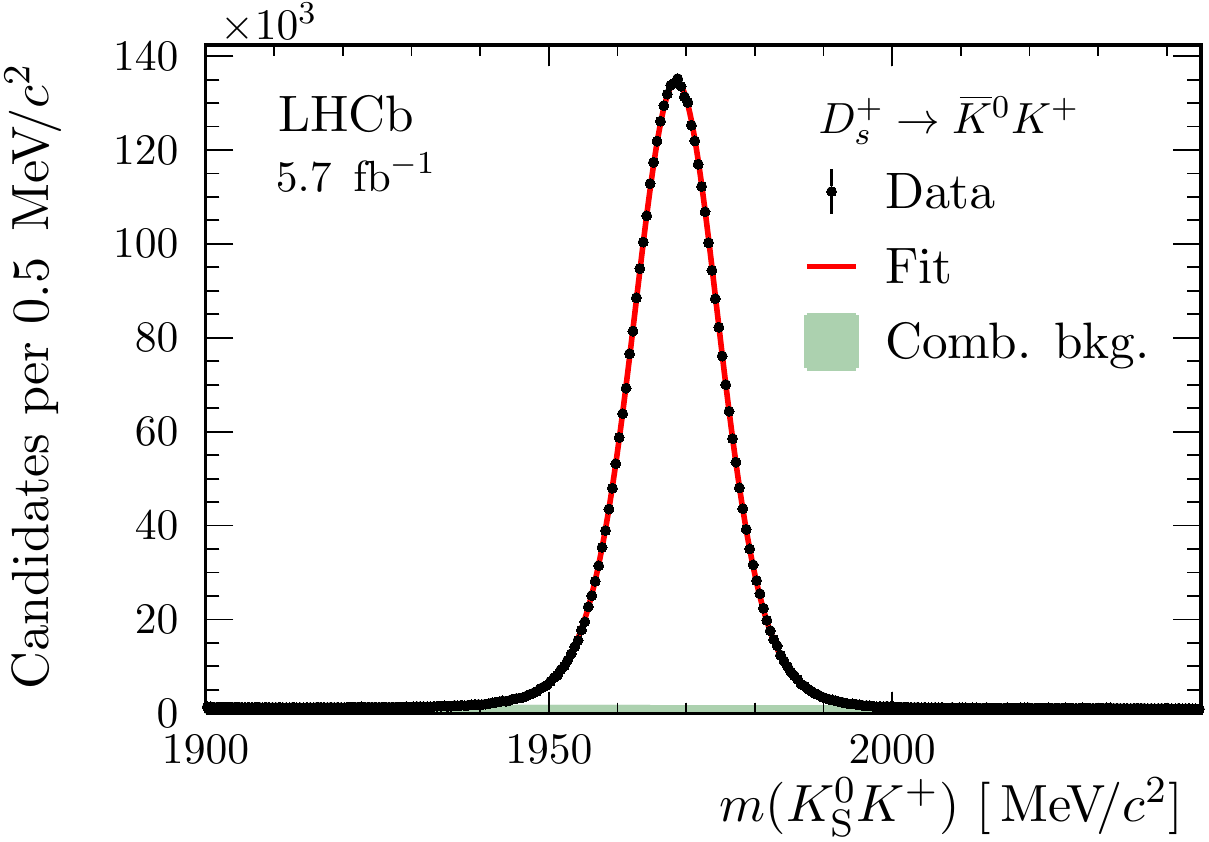}\\
\caption{Distributions of the invariant mass for the weighted charm-meson candidates, for
the decays (top left) \Dstarp\to\Dz(\to\Km\Kp)\pip, (top right) \Dstarp\to\Dz(\to\Km\pip)\pip, (bottom left) \Dsp\to\Km\Kp\pip, and
(bottom right) \Dsp\to\Kzb\Kp. The data are from the
\CDS calibration procedure. The results of the fits to these
distributions are also shown.}
\end{figure}

\begin{figure}[htb!]
\centering
\includegraphics[width=0.49\textwidth]{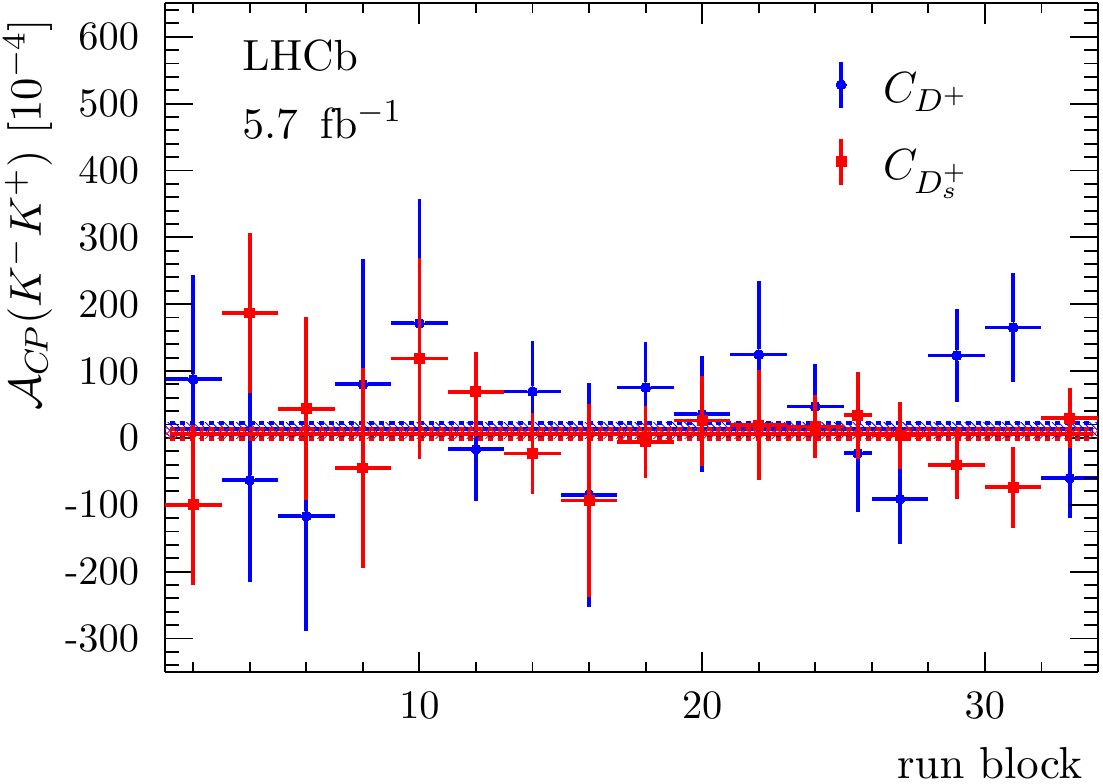}
\includegraphics[width=0.49\textwidth]{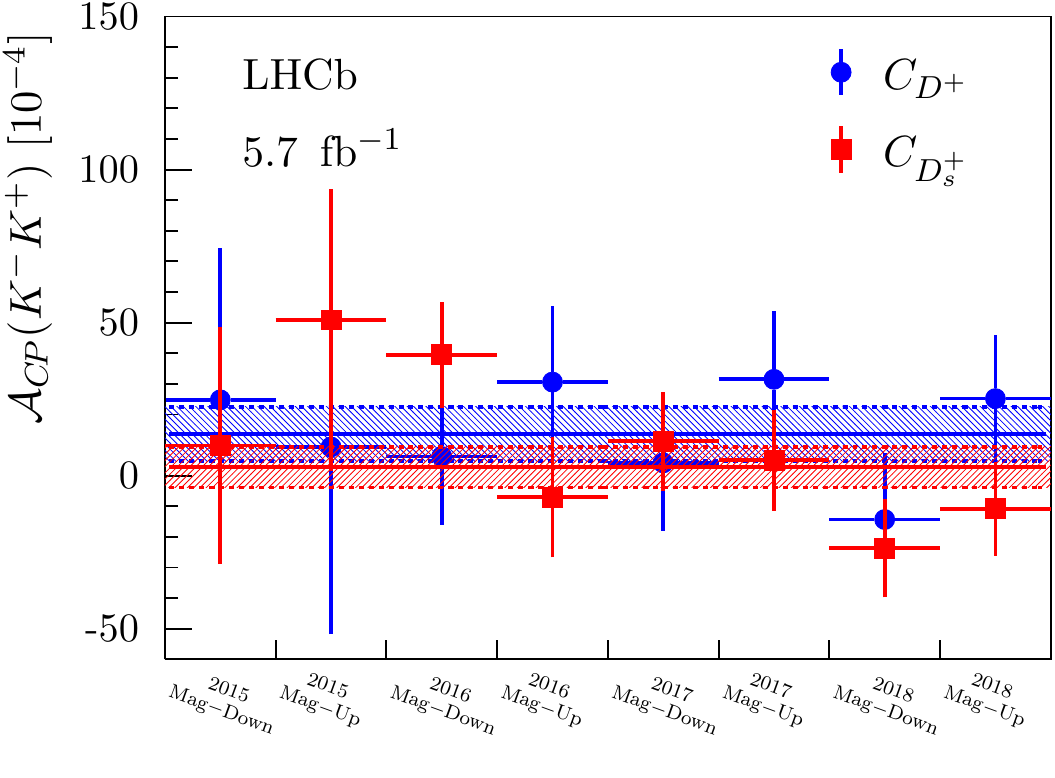}
\caption{Measurements of $\ACPKK$ in (left) time-ordered data-taking (referred to as run blocks) and (right) year of data-taking and dipole-magnet polarity for the \CDP and \CDS calibration procedures. The uncertainties are statistical only. The horizontal lines are the averaged values for the \CDP and \CDS methods, while the bands represent the one-standard-deviation regions. The labels \textit{Mag-Up} and \textit{Mag-Down} refer to the direction of the magnetic field along the positive and negative directions of $y$-axis, respectively. } 
\end{figure}

\begin{figure}[htb!]
\centering
\includegraphics[width=0.49\textwidth]{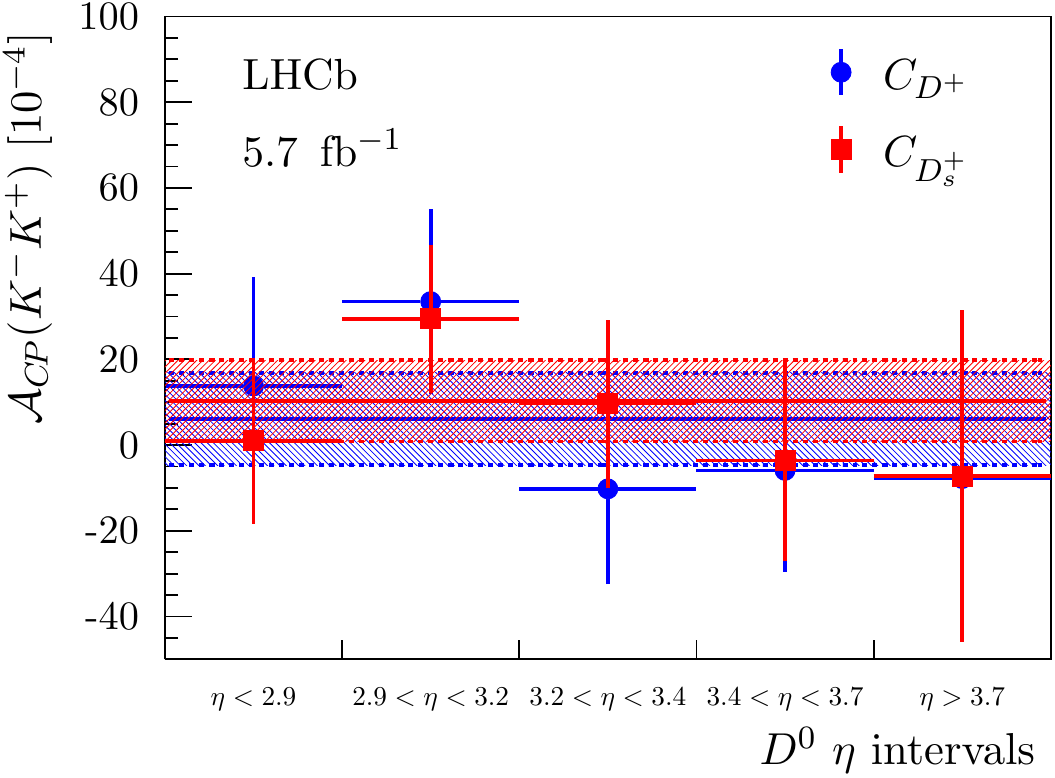}
\includegraphics[width=0.49\textwidth]{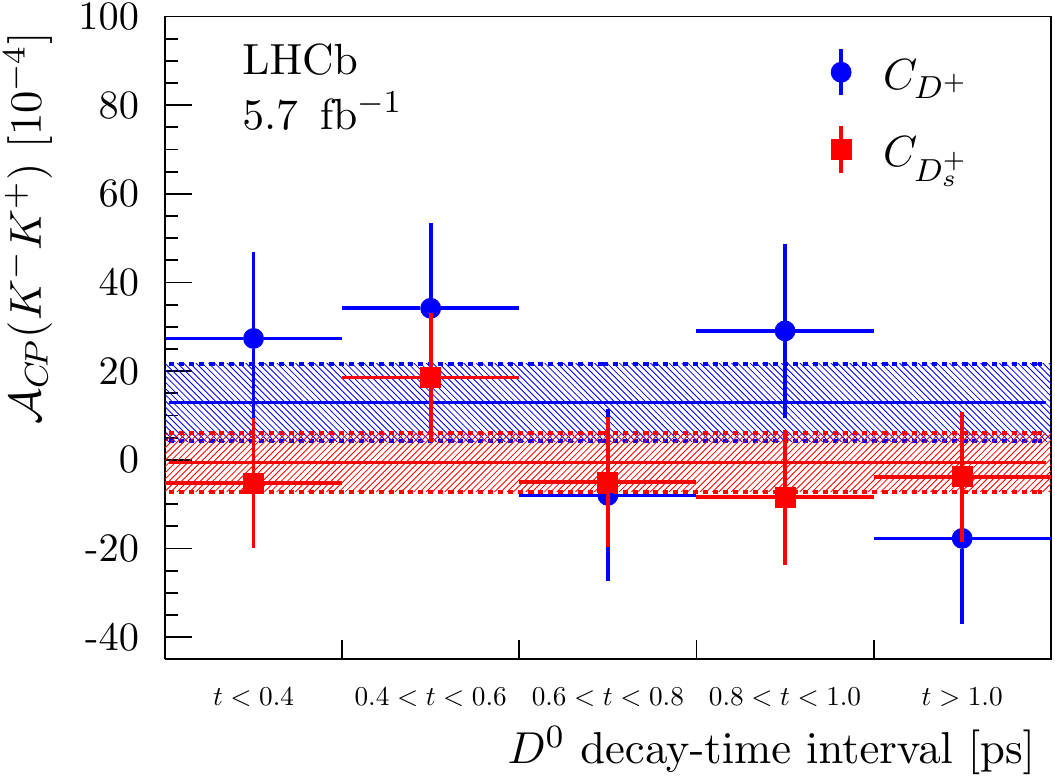}
\caption{Comparison of the results for $\ACPKK$ per different intervals of \Dz candidates (left) pseudorapidity and (right) decay time for the \CDP and \CDS calibration procedures. The horizontal lines are the averaged values for the \CDP and \CDS methods, while the bands represent the one-standard-deviation regions.}
\end{figure}

\begin{figure}[htb!]
\centering
\includegraphics[width=0.80\textwidth]{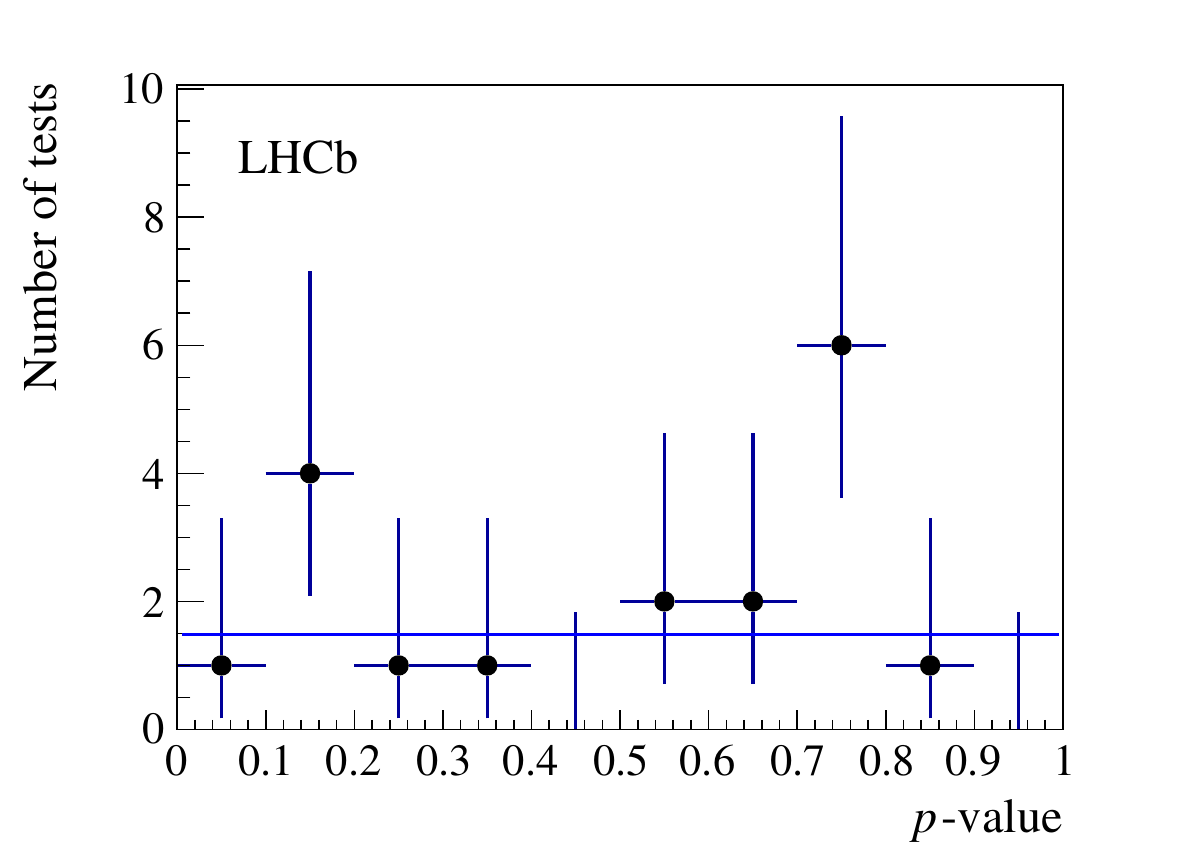}
\caption{Distribution of the $p$-values resulting from the consistency checks on \ACPKK measured in different data subsamples. The blue line represents the expected uniform distribution.}
\end{figure}

\begin{figure}[htb!]
\centering
\includegraphics[width=0.80\textwidth]{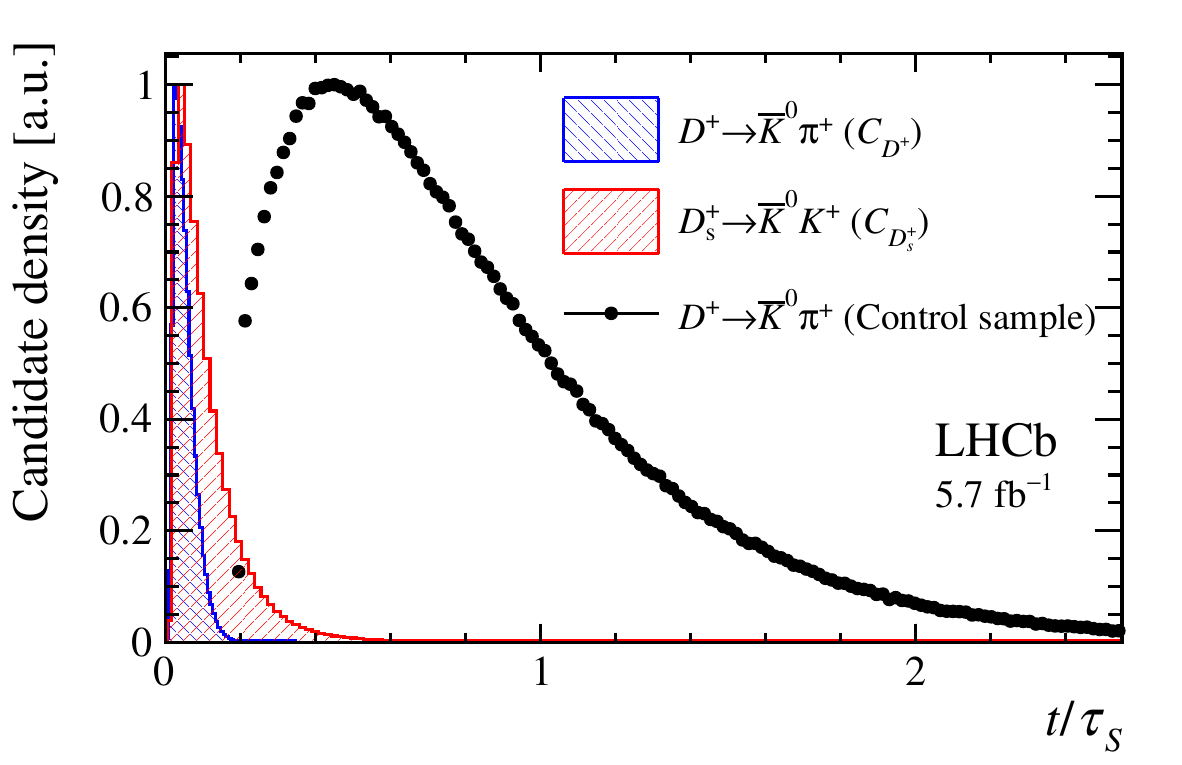}
\caption{Distributions of the neutral-kaon decay time for background-subtracted and weighted candidates considered in the determination of $\acpkk$ values. The distributions are normalized to have the maximum equal to unity. The control sample is used to estimate $A(\Kzb)$ and consists of neutral kaons decaying outside of the vertex detector.}
\end{figure}

\begin{figure}[htb!]
\centering
\includegraphics[width=0.80\textwidth]{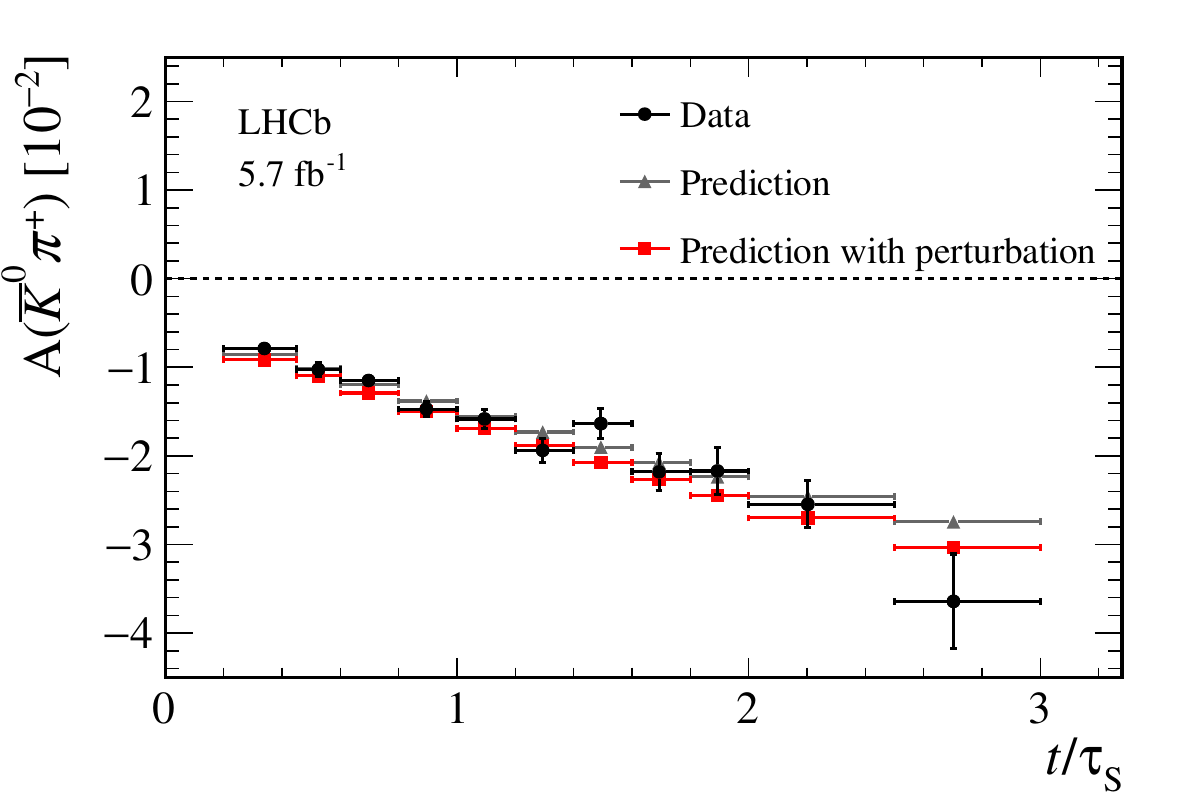}
\caption{Measured asymmetry in \Dp\to\Kzb\pip decays as a function of \Kzb decay time in units of \KS-meson decay time, in data where the neutral kaon meson decays outside of the vertex detector. The predictions for the neutral kaon asymmetry, from the model with and without an additional linear degree of freedom, are shown as well. An overall shift is applied to the model's predictions to account for unrelated nuisance asymmetries, \ie $\Aprod(\Dp)$ and $\Adet(\pip)$.}
\end{figure}

\begin{figure}[htb!]
\centering
\includegraphics[width=0.80\textwidth]{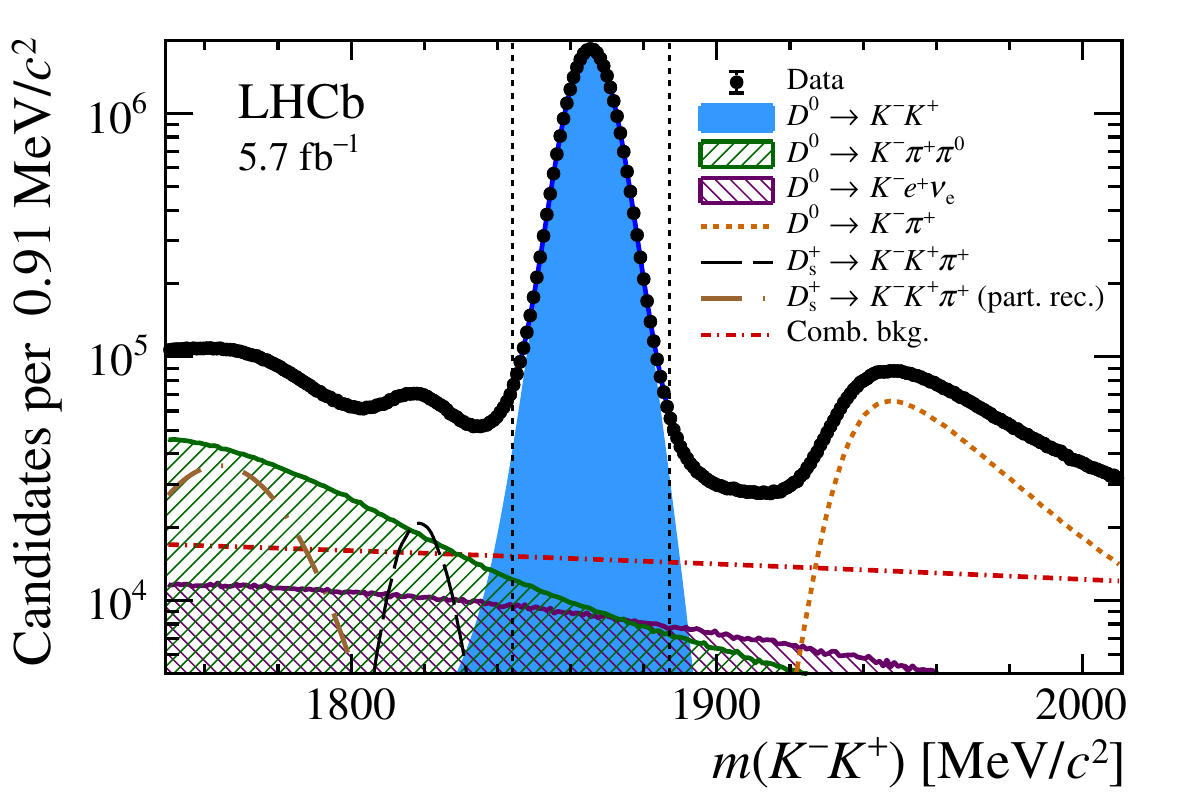}
\caption{Distribution of the $\Km\Kp$ invariant mass, along with the result of the fit describing the expected background components. The vertical lines indicate the interval applied in the selection of \Dz candidates for the determination of $\acpkk$. The fit results are used to evaluate the systematic uncertainties related to the peaking background.}
\end{figure}

\begin{figure}[htb!]
\centering
\includegraphics[width=0.49\textwidth]{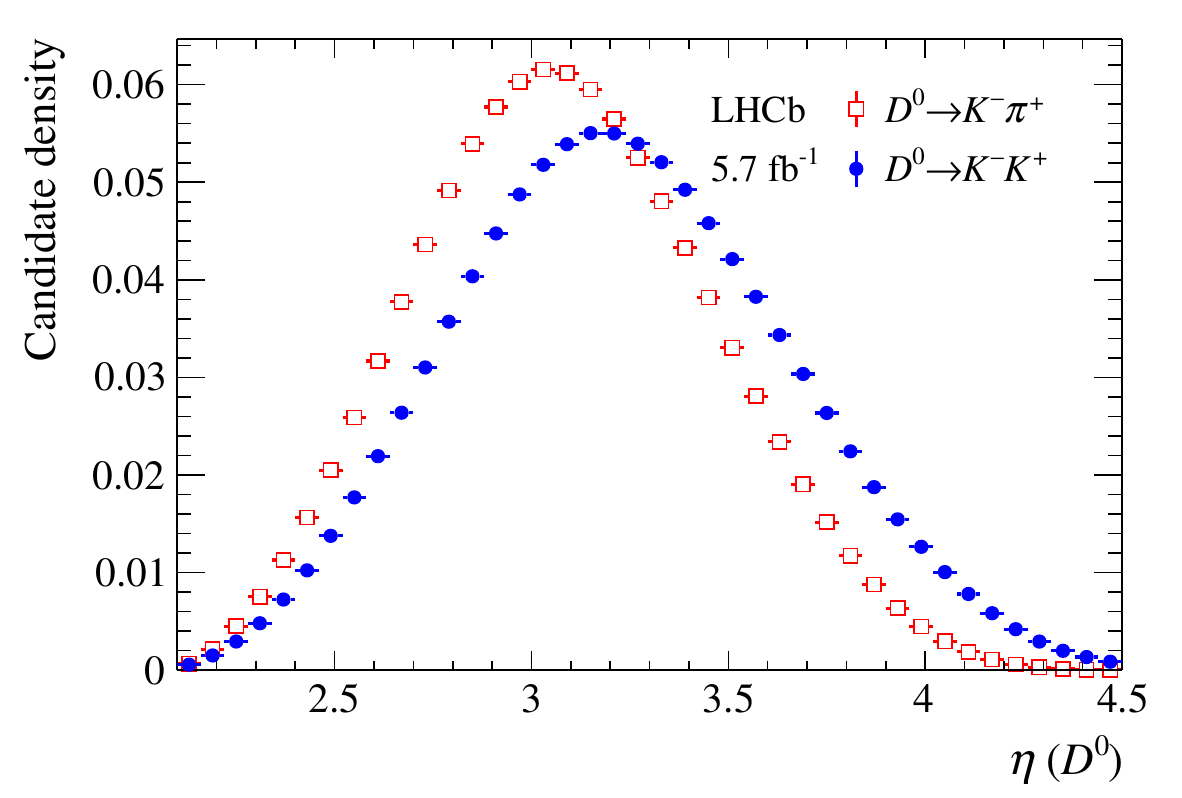}
\includegraphics[width=0.49\textwidth]{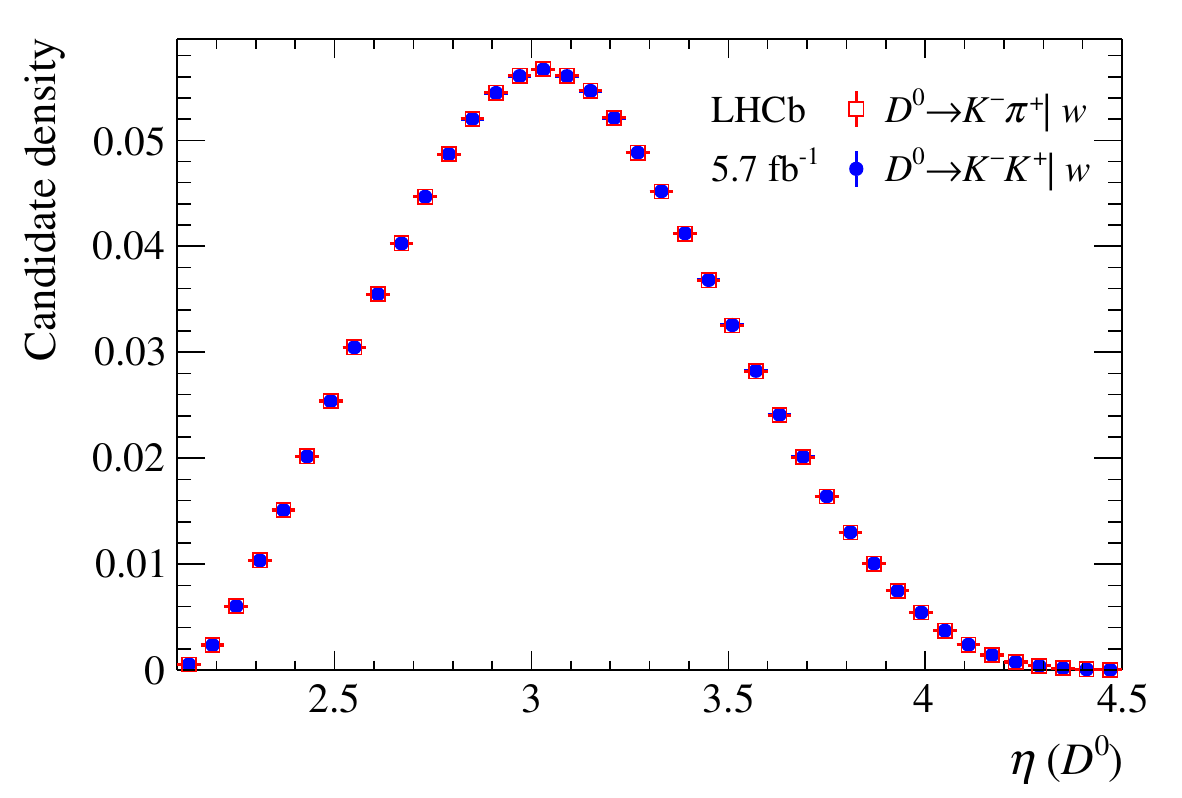} \\
\includegraphics[width=0.49\textwidth]{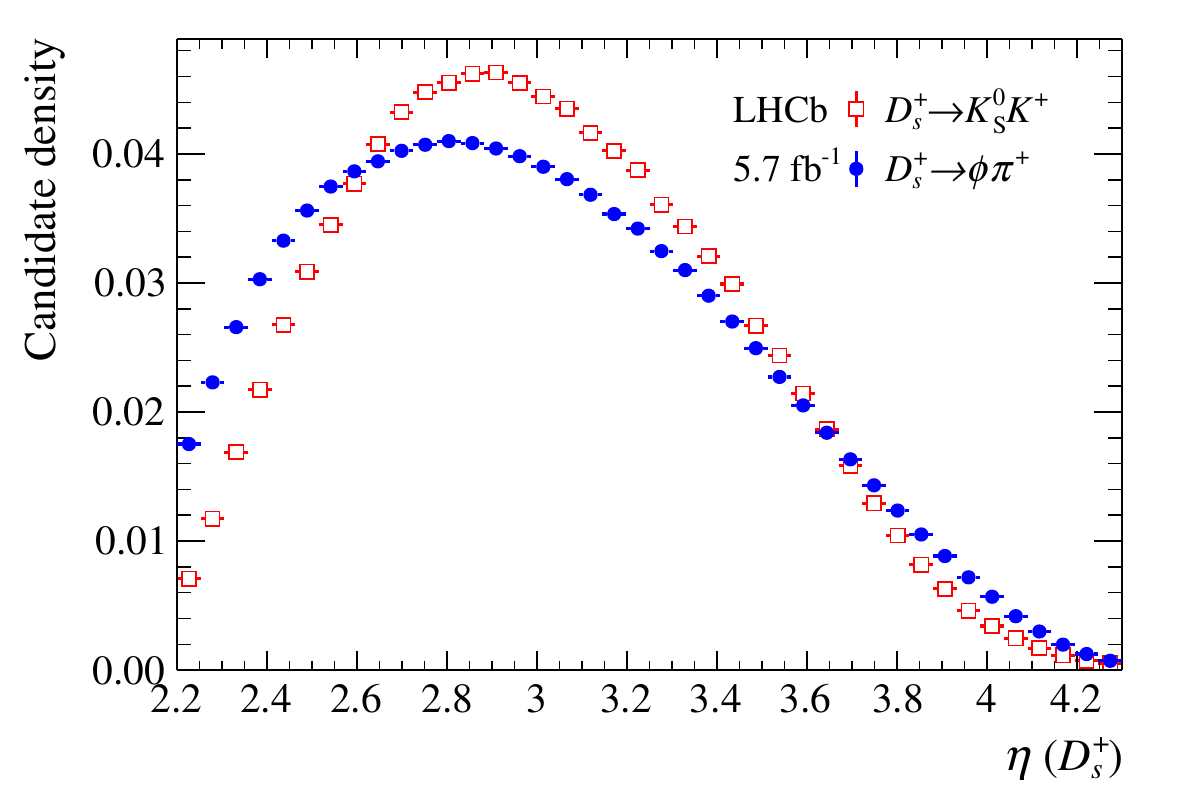}
\includegraphics[width=0.49\textwidth]{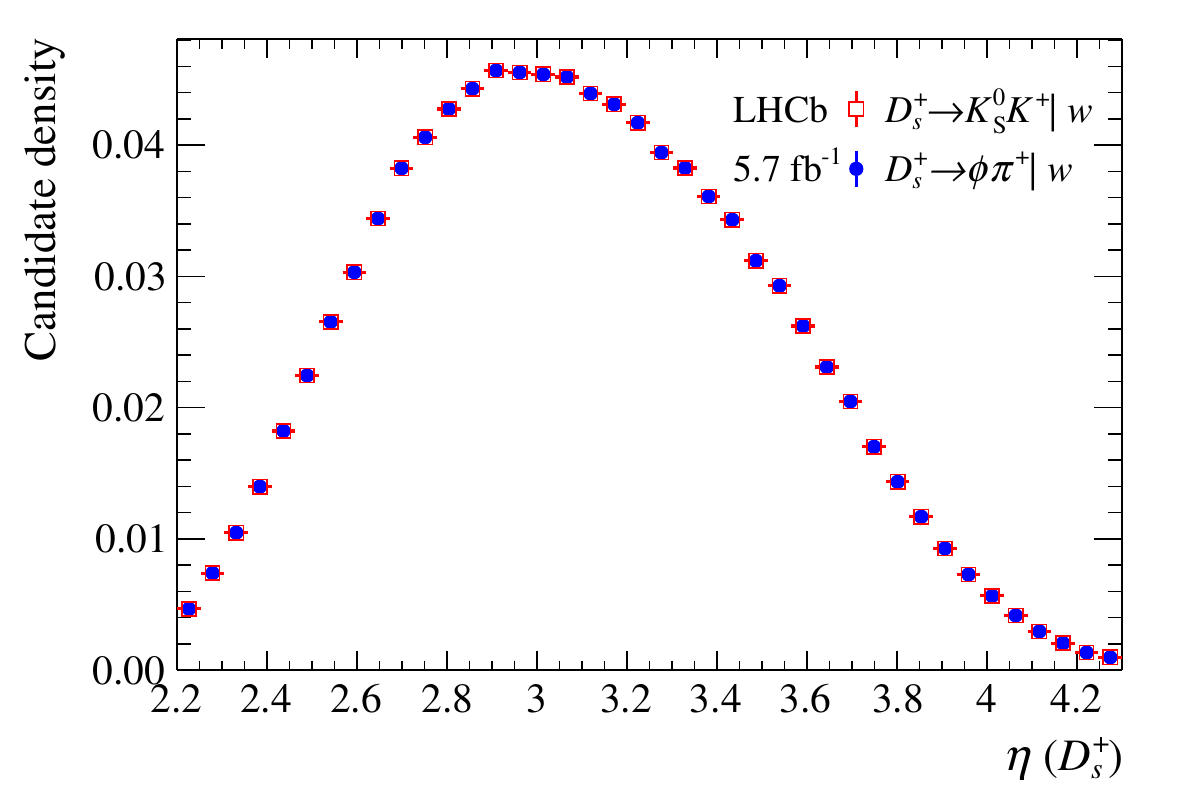} \\
\includegraphics[width=0.49\textwidth]{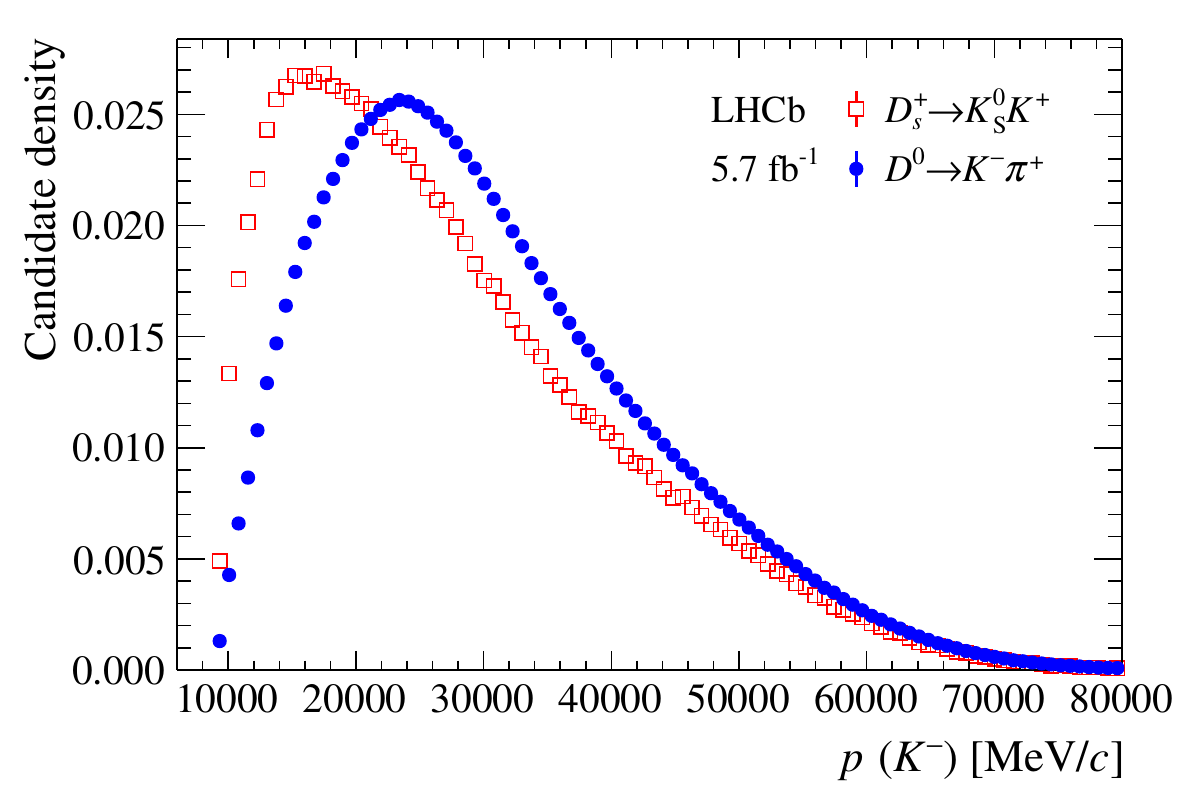}
\includegraphics[width=0.49\textwidth]{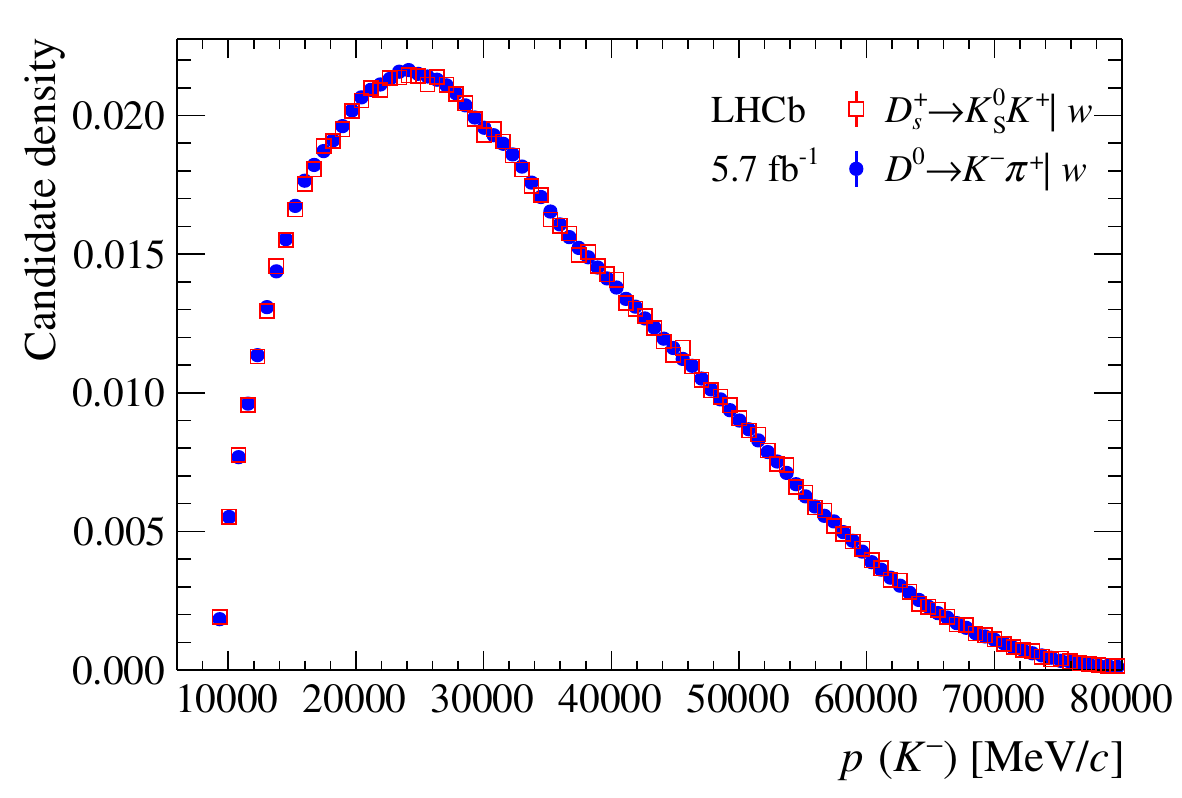}\\
\includegraphics[width=0.49\textwidth]{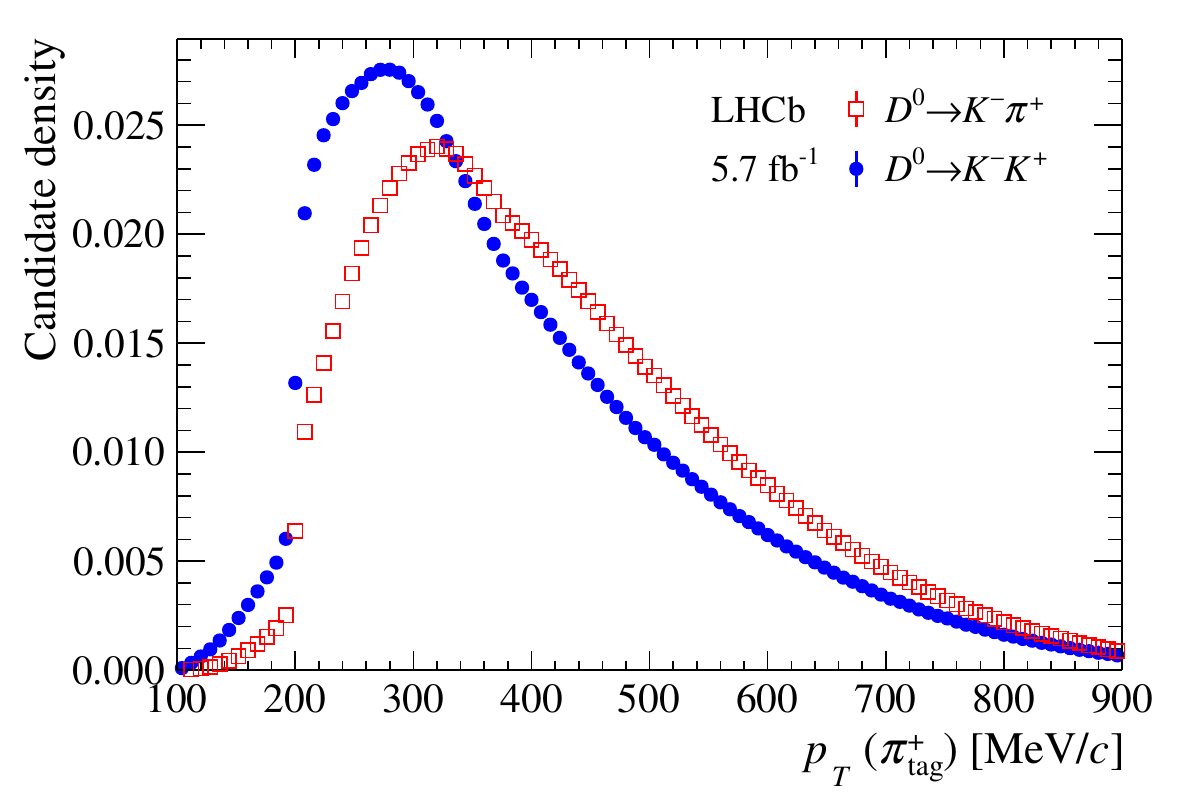}
\includegraphics[width=0.49\textwidth]{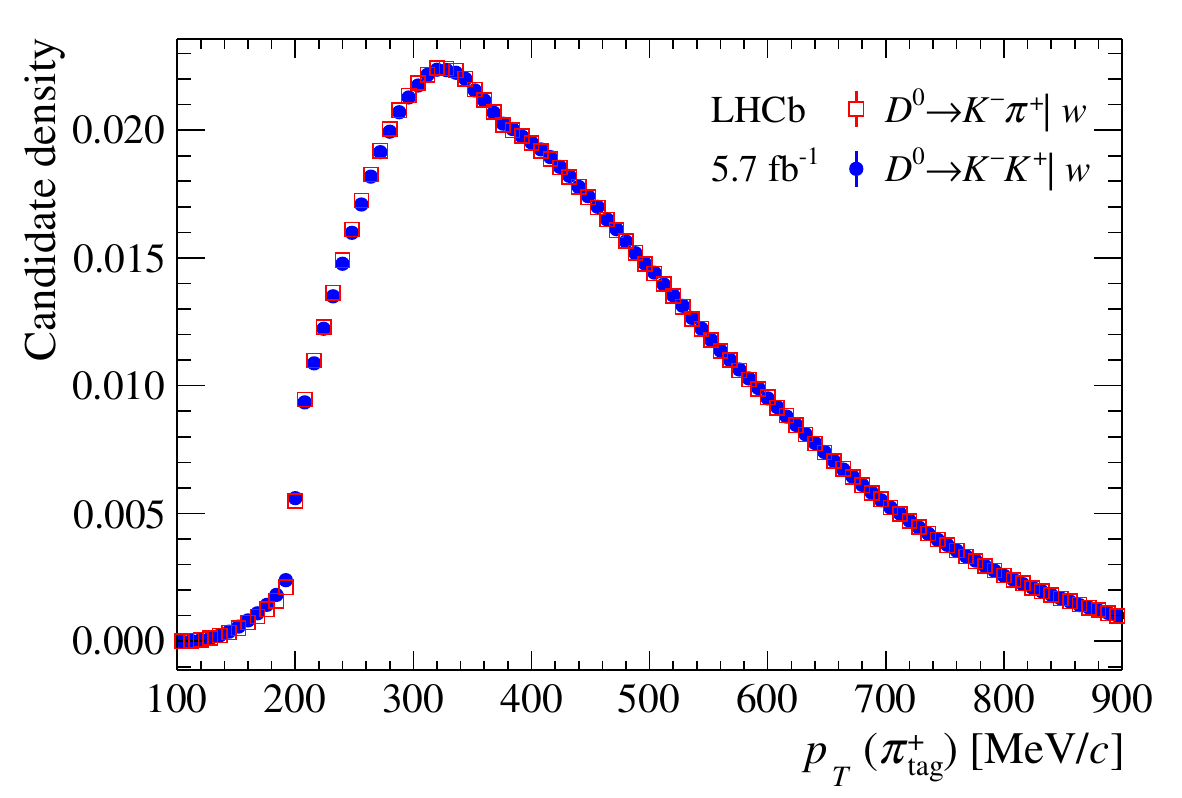}
\caption{Background-subtracted kinematic distributions for the charm-meson pseudorapidity ($\eta$), the kaon momentum ($p$) and the tagging-pion transverse momentum (\pt) (left column) before and (right column) after the per-candidate \CDS weighting procedure. The distributions are normalized to unit area.}
\end{figure}

\begin{figure}[htb!]
\centering
\includegraphics[width=0.49\textwidth]{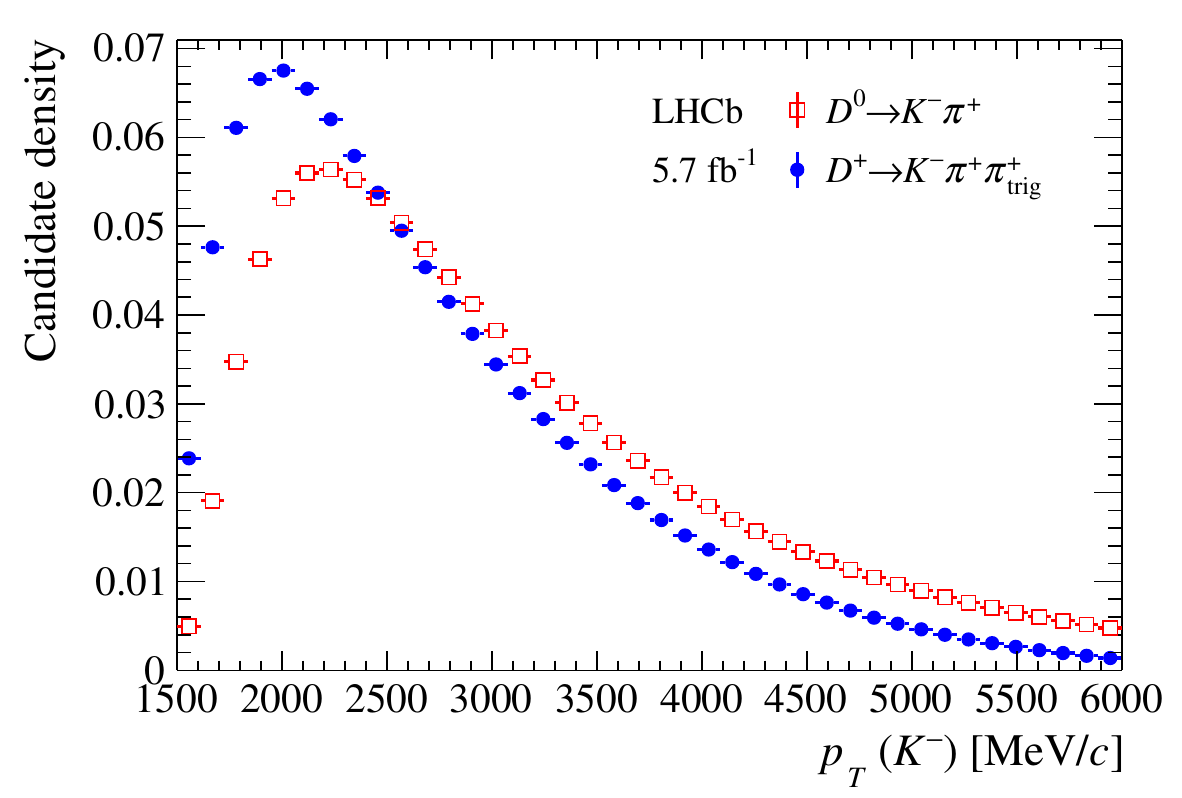}
\includegraphics[width=0.49\textwidth]{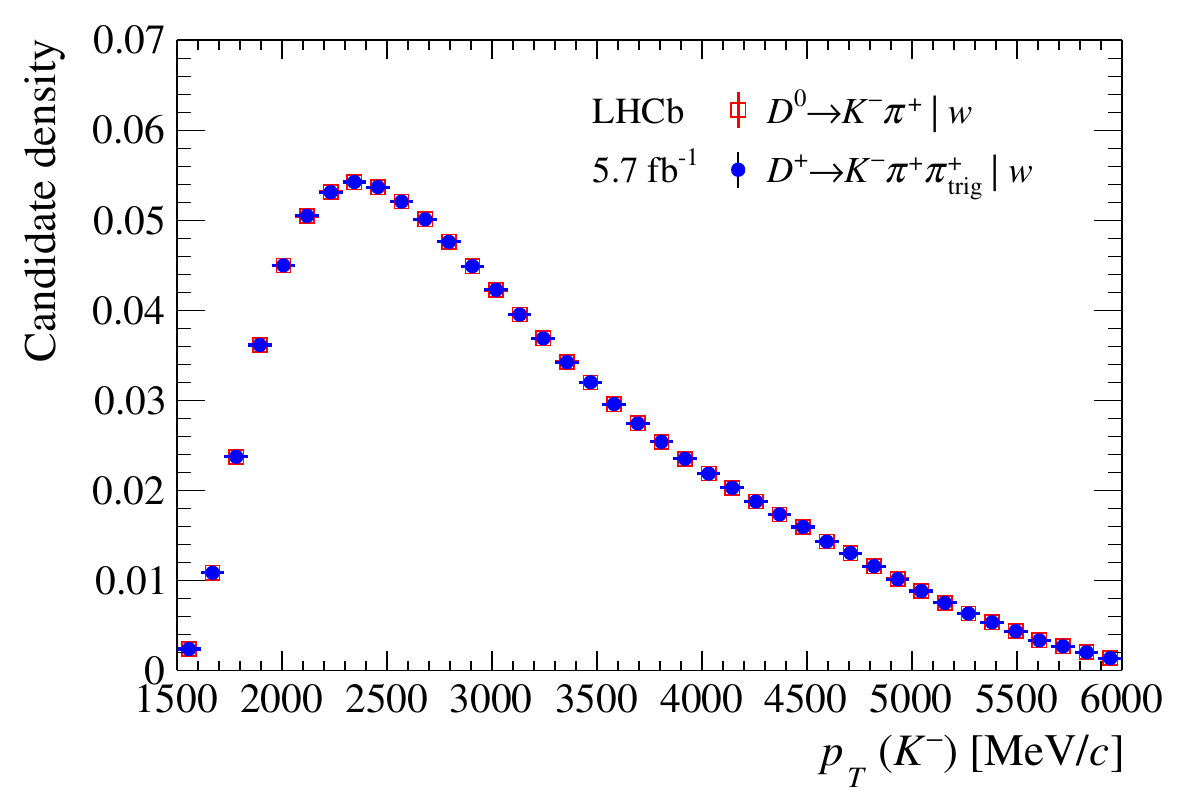} \\
\includegraphics[width=0.49\textwidth]{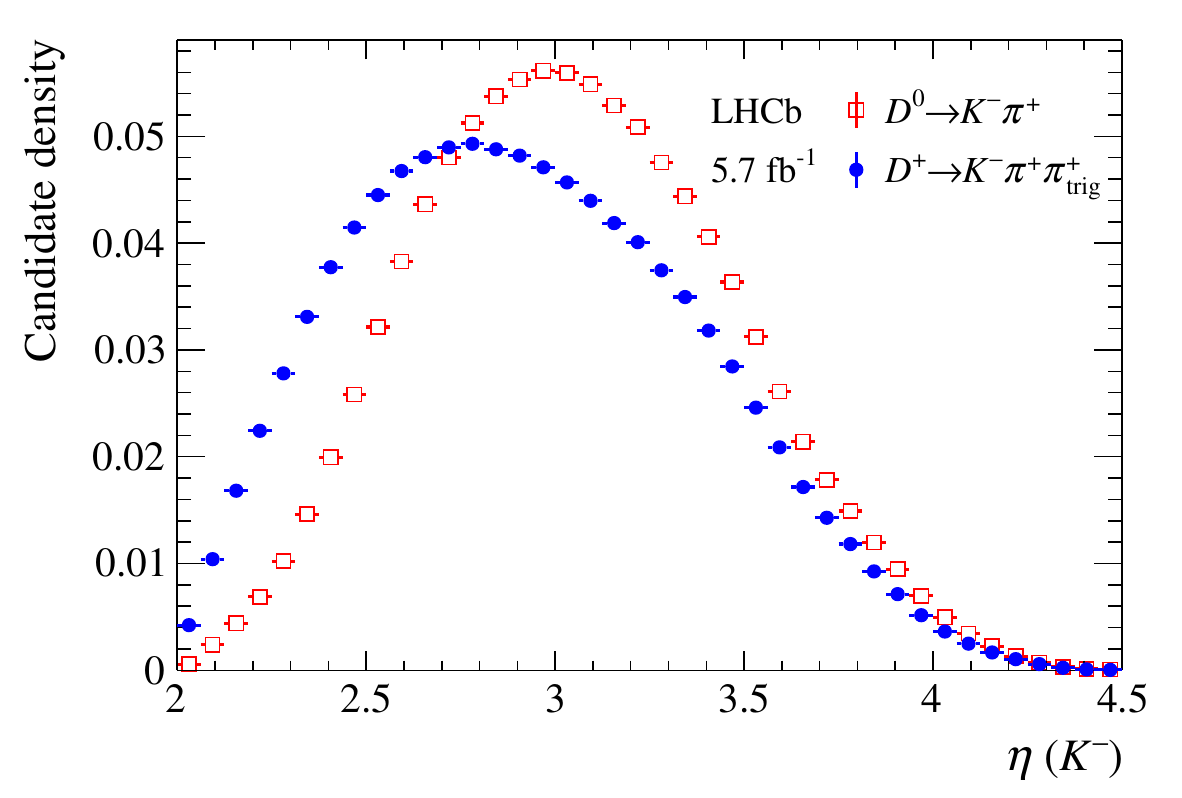}
\includegraphics[width=0.49\textwidth]{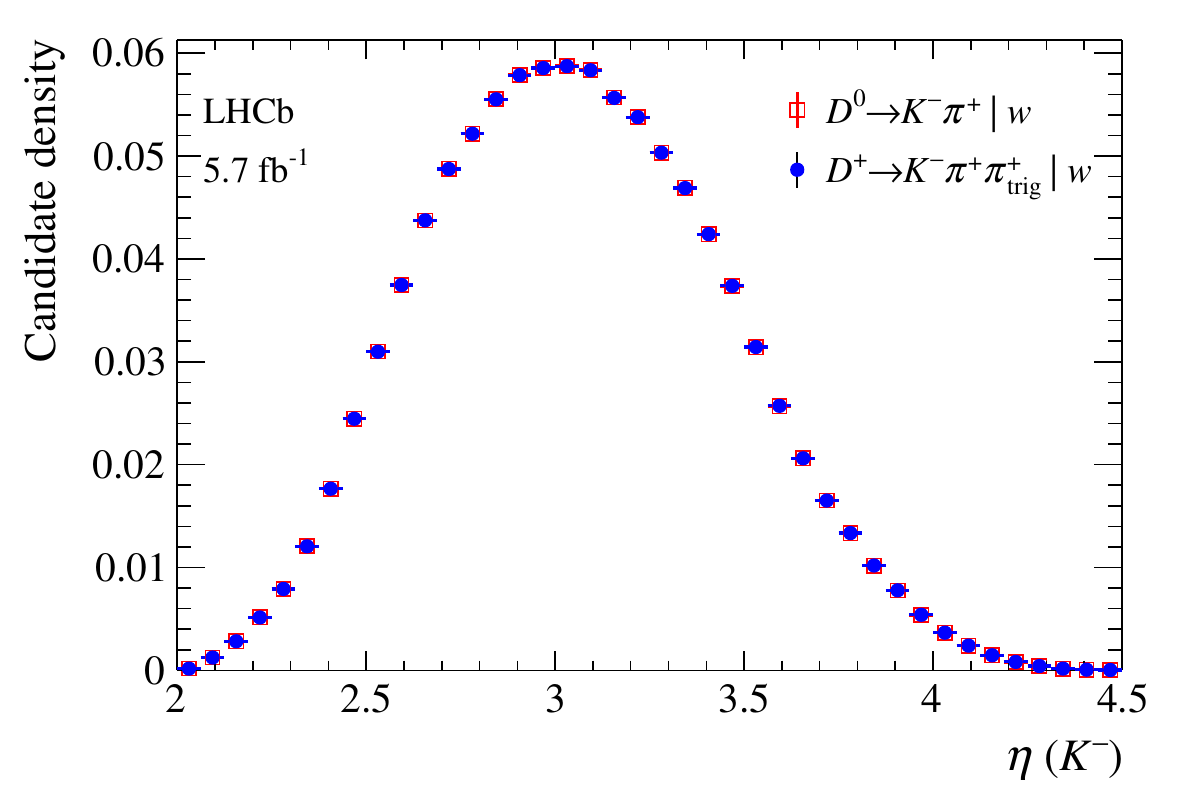} \\
\includegraphics[width=0.49\textwidth]{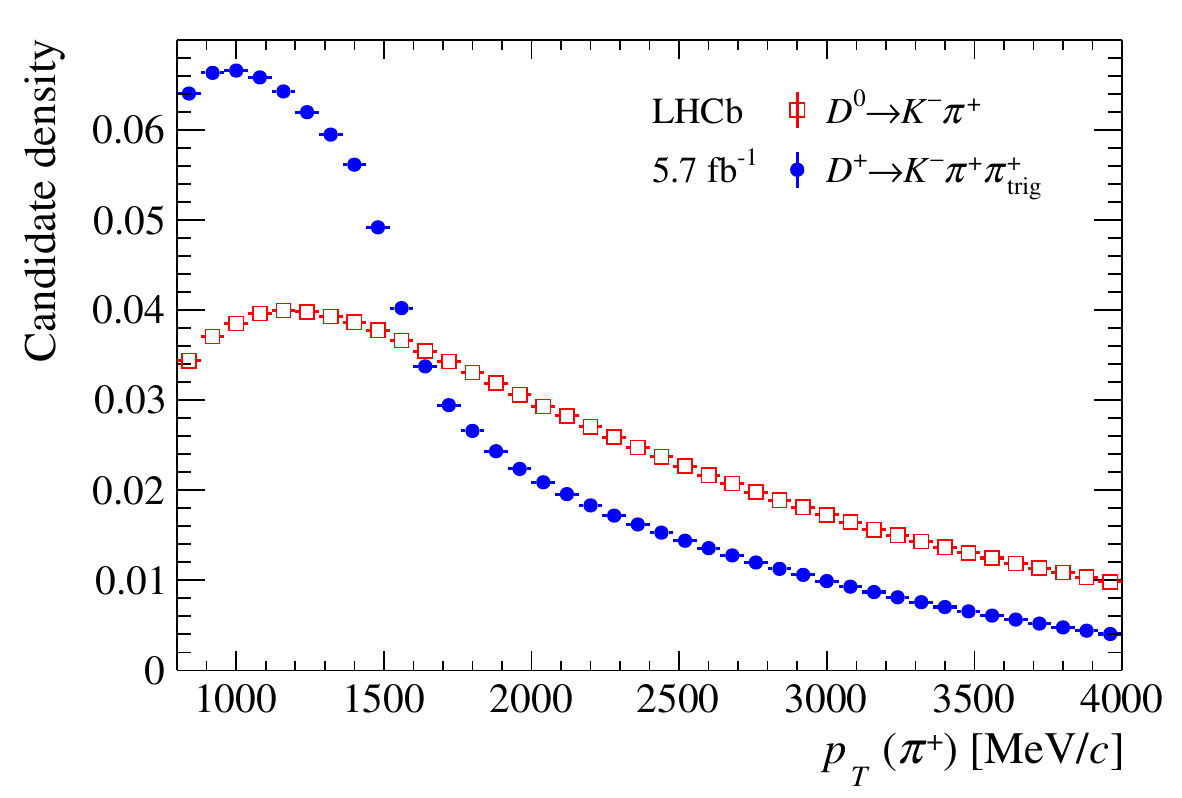}
\includegraphics[width=0.49\textwidth]{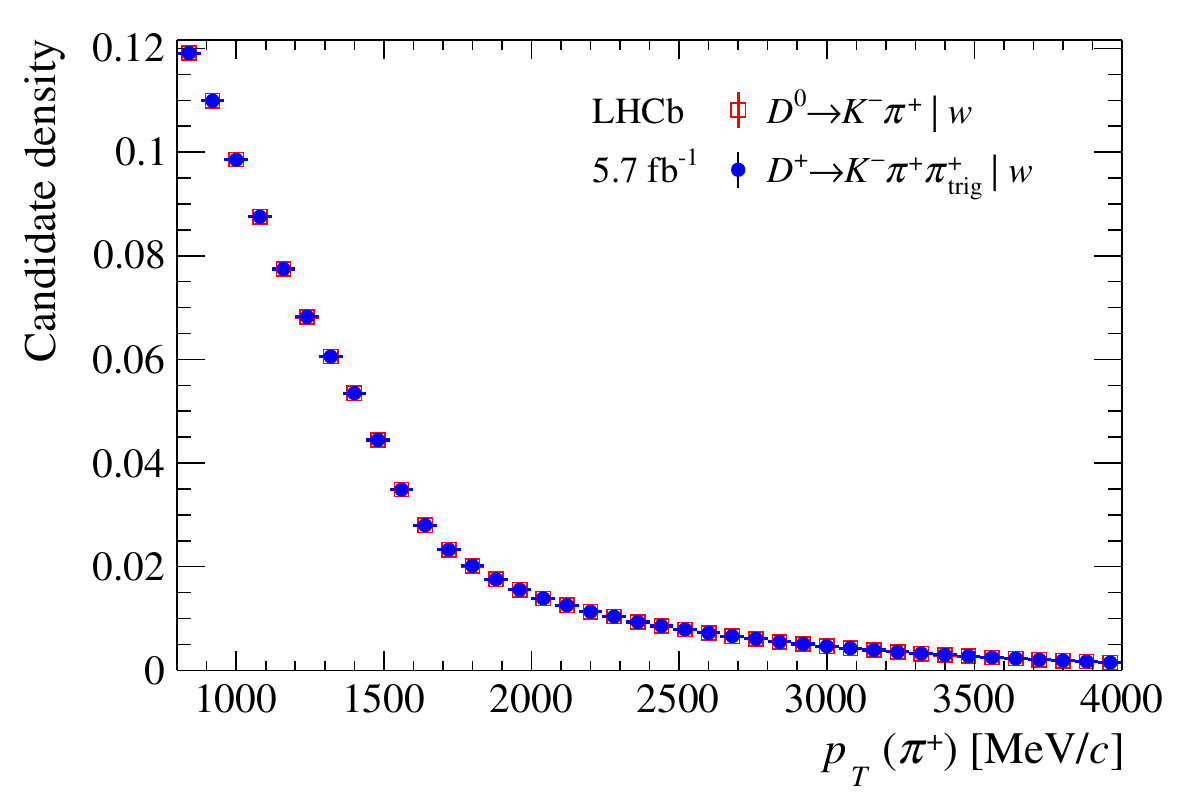}\\
\includegraphics[width=0.49\textwidth]{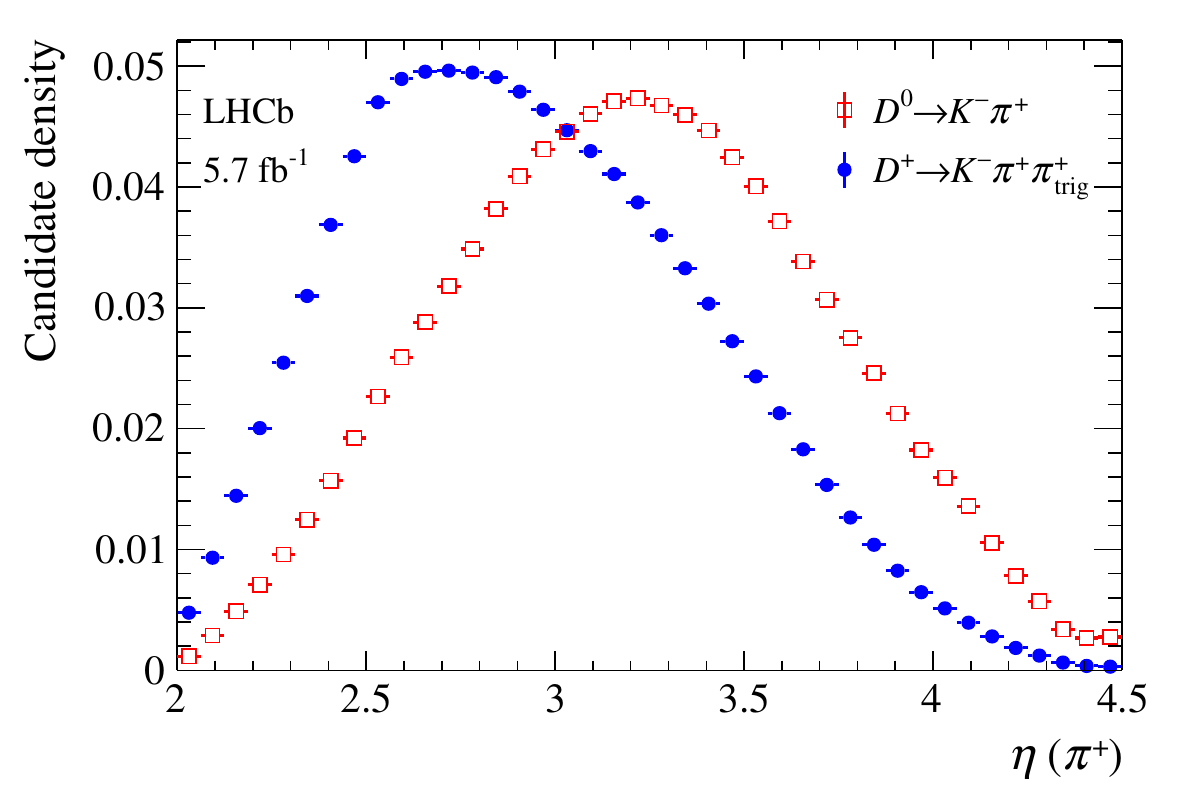}
\includegraphics[width=0.49\textwidth]{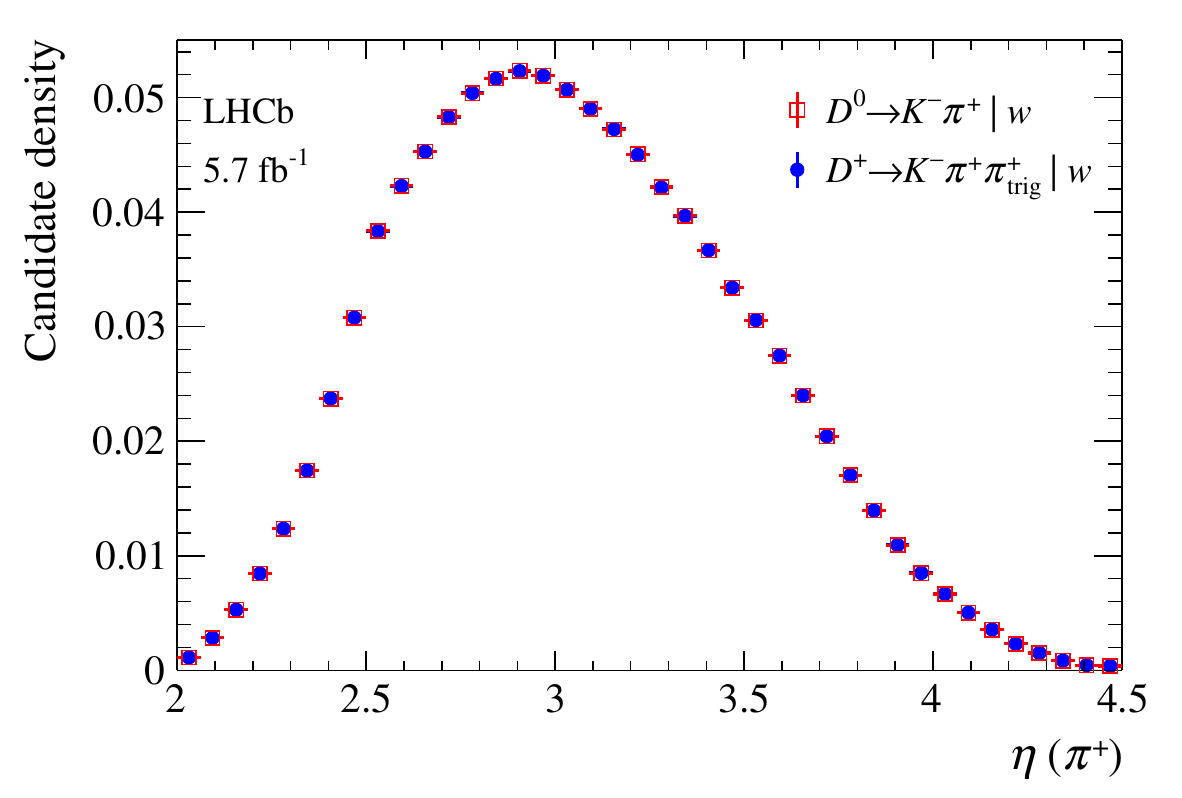}\\
\caption{Background-subtracted kinematic distributions for the kaon and pion transverse momentum ($\pt$) and pseudorapidity ($\eta$) (left column) before and (right column) after the per-candidate \CDP weighting procedure. The distributions are normalized to unit area.}
\end{figure}

\begin{figure}[htb!]
\centering
\includegraphics[width=0.80\textwidth]{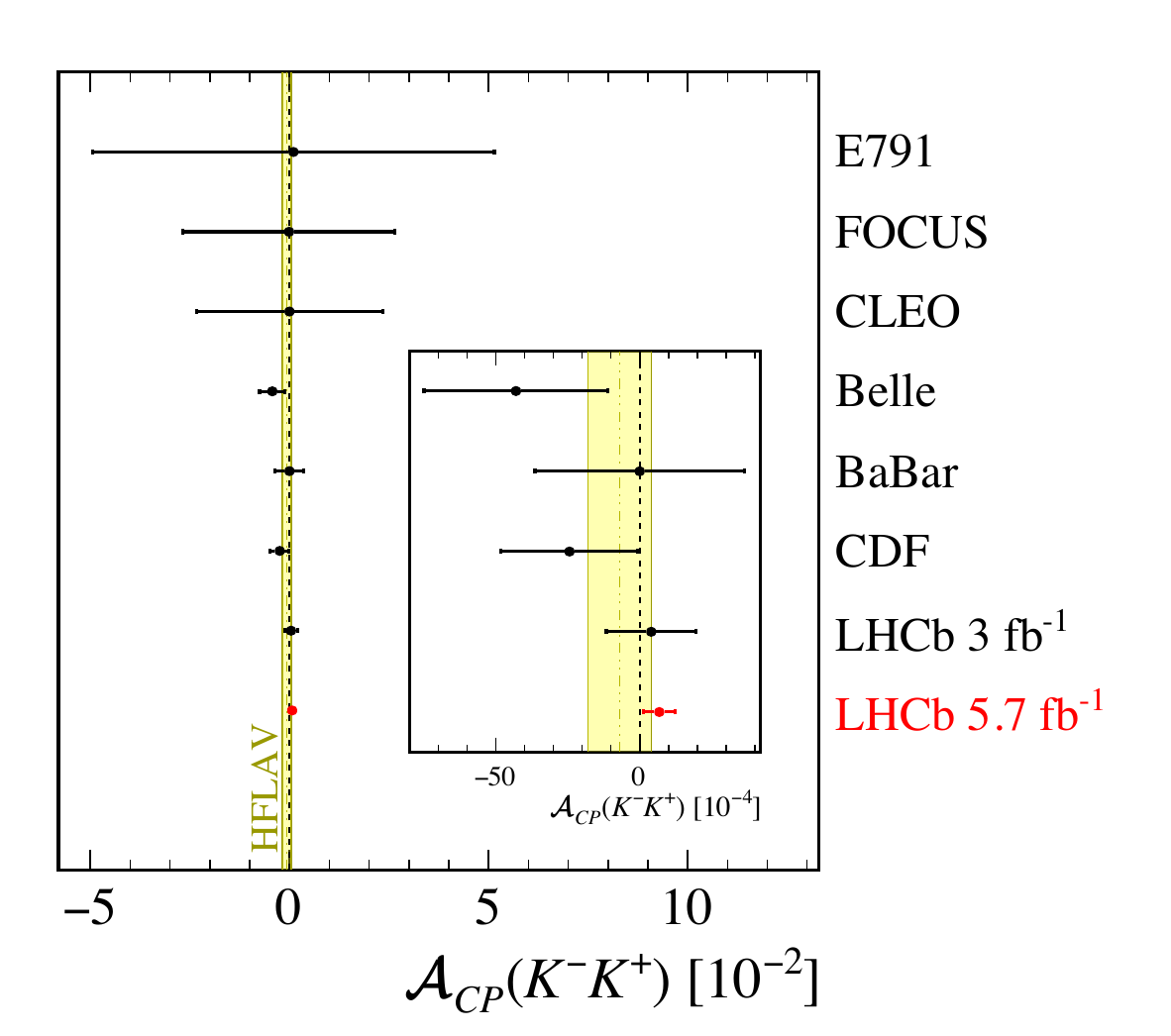}
\caption{Measurements of \ACPKK from various experiments~\cite{E791:1997txw,FOCUS:2000ejh,CLEO:2001lgl,Belle:2008ddg,BaBar:2007tfw,CDF:2011ejf,LHCb-PAPER-2016-035}. The presented measurement is highlighted in red. The vertical band corresponds to the average of all measurements previous to the presented, computed by HFLAV~\cite{HFLAV18}, where it is assumed that \CP is conserved in mixing and in the interference between decay and mixing. This assumption is necessary when results from different experiments are compared. The inset plot shows the five most precise measurements in a reduced horizontal range. }
\end{figure}

\clearpage 
\addcontentsline{toc}{section}{References}
\bibliographystyle{LHCb}
\bibliography{main,standard,LHCb-PAPER,LHCb-CONF,LHCb-DP,LHCb-TDR}

\newpage
\centerline
{\large\bf LHCb collaboration}
\begin
{flushleft}
\small
R.~Aaij$^{32}$\lhcborcid{0000-0003-0533-1952},
A.S.W.~Abdelmotteleb$^{50}$\lhcborcid{0000-0001-7905-0542},
C.~Abellan~Beteta$^{44}$,
F.~Abudin{\'e}n$^{50}$\lhcborcid{0000-0002-6737-3528},
T.~Ackernley$^{54}$\lhcborcid{0000-0002-5951-3498},
B.~Adeva$^{40}$\lhcborcid{0000-0001-9756-3712},
M.~Adinolfi$^{48}$\lhcborcid{0000-0002-1326-1264},
P.~Adlarson$^{77}$\lhcborcid{0000-0001-6280-3851},
H.~Afsharnia$^{9}$,
C.~Agapopoulou$^{13}$\lhcborcid{0000-0002-2368-0147}
% ...

{\footnotesize \it

% ...
$^{9}$Universit{\'e} Clermont Auvergne, CNRS/IN2P3, LPC, Clermont-Ferrand, France\\
% ...
$^{13}$LPNHE, Sorbonne Universit{\'e}, Paris Diderot Sorbonne Paris Cit{\'e}, CNRS/IN2P3, Paris, France\\
% ...
$^{32}$Nikhef National Institute for Subatomic Physics, Amsterdam, Netherlands\\
% ...
$^{40}$Instituto Galego de F{\'\i}sica de Altas Enerx{\'\i}as (IGFAE), Universidade de Santiago de Compostela, Santiago de Compostela, Spain\\
% ...
$^{44}$Physik-Institut, Universit{\"a}t Z{\"u}rich, Z{\"u}rich, Switzerland\\
% ...
$^{48}$H.H. Wills Physics Laboratory, University of Bristol, Bristol, United Kingdom\\
% ...
$^{50}$Department of Physics, University of Warwick, Coventry, United Kingdom\\
% ...
$^{53}$School of Physics and Astronomy, University of Glasgow, Glasgow, United Kingdom\\
$^{54}$Oliver Lodge Laboratory, University of Liverpool, Liverpool, United Kingdom\\
% ...
$^{77}$Department of Physics and Astronomy, Uppsala University, Uppsala, Sweden, associated to $^{53}$\\
% ...
}
\end{flushleft}
\end{document}